\documentclass[a4paper,11pt]{article}
\usepackage{jheppub}

\usepackage{chessfss}
\pdfoutput=1
\pdfpageattr {/Group << /S /Transparency /I true /CS /DeviceRGB>>}
\usepackage{hyperref}
\hypersetup{colorlinks=true,linkcolor=blue,citecolor=blue}
\usepackage{amsmath,amssymb,amsfonts}
\usepackage{bm}
\usepackage{float}
\usepackage{subfig}
\usepackage{scalefnt}
\usepackage{wrapfig}
\usepackage{fancybox}
\usepackage{tikz}
\usepackage{multirow}

\newcommand{\eq}[1]{Eq.~(\ref{#1})} 
\newcommand{\cmt}[1]{\textcolor{red}{\textbf{(#1)}}} 
\allowdisplaybreaks

\usetikzlibrary{arrows}
\tikzset{>=angle 60}
\tikzstyle{W}=[draw,circle,scale=.6]
\tikzstyle{B}=[draw,circle,fill=black,scale=.6]
\tikzstyle{H}=[draw,circle,fill=gray,scale=.6]
\tikzstyle{every picture}=[scale=.6,baseline=(current bounding box.south)]

\def\a{\alpha}
\def\b{\beta}
\def\c{\varepsilon}
\def\d{\delta}
\def\e{\epsilon}
\def\f{\phi}
\def\g{\gamma}
\def\h{\theta}
\def\k{\kappa}
\def\l{\lambda}
\def\m{\mu}
\def\n{\nu}
\def\p{\psi}
\def\q{\partial}
\def\r{\rho}
\def\s{\sigma}
\def\t{\tau}
\def\u{\upsilon}
\def\v{\varphi}
\def\w{\omega}
\def\x{\xi}
\def\y{\eta}
\def\z{\zeta}
\def\D{\Delta}
\def\G{\Gamma}
\def\H{\Theta}
\def\L{\zeta}
\def\F{\Phi}
\def\P{\Psi}
\def\S{\Sigma}

\def\aa{{\dot \a}}
\def\bb{{\dot \b}}
\def\ss{{\bar \s}}
\def\hh{{\bar \h}}
\def\CA{{\cal A}}
\def\CB{{\cal B}}
\def\CC{{\cal C}}
\def\CD{{\cal D}}
\def\CE{{\cal E}}
\def\CG{{\cal G}}
\def\CH{{\cal H}}
\def\CI{{\cal I}}
\def\CK{{\cal K}}
\def\CL{{\cal L}}
\def\CR{{\cal R}}
\def\CM{{\cal M}}
\def\CN{{\cal N}}
\def\CO{{\cal O}}
\def\CP{{\cal P}}
\def\CQ{{\cal Q}}
\def\CW{{\cal W}}

\DeclareMathOperator{\Tr}{Tr}
\newcommand{\Slash}[1]{{\ooalign{\hfil/\hfil\crcr$#1$}}}

\def\o{\over}
\newcommand{\gsim}{ \mathop{}_{\textstyle \sim}^{\textstyle >} }
\newcommand{\lsim}{ \mathop{}_{\textstyle \sim}^{\textstyle <} }
\newcommand{\vev}[1]{ \left\langle {#1} \right\rangle }
\newcommand{\EV}{ {\rm eV} }
\newcommand{\KEV}{ {\rm keV} }
\newcommand{\MEV}{ {\rm MeV} }
\newcommand{\GEV}{ {\rm GeV} }
\newcommand{\TEV}{ {\rm TeV} }
\def\diag{\mathop{\rm diag}\nolimits}
\def\Spin{\mathop{\rm Spin}}
\def\SO{\mathop{\rm SO}}
\def\O{\mathop{\rm O}}
\def\SU{\mathop{\rm SU}}
\def\U{\mathrm{U}}
\def\Sp{\mathop{\rm Sp}}
\def\SL{\mathop{\rm SL}}
\def\tr{\mathop{\rm tr}}
\def\rank{\mathop{\rm rank}}

\def\beq#1\eeq{\begin{align}#1\end{align}}

\newcommand{\be}{\begin{eqnarray}}
\newcommand{\ee}{\end{eqnarray}}
\newcommand{\bea}{\begin{eqnarray}}
\newcommand{\eea}{\end{eqnarray}}

\newcommand{\nn}{\nonumber}
\newcommand{\bn}{\begin{enumerate}}
\newcommand{\en}{\end{enumerate}}

\newcommand{\ada}[4]{\ensuremath{\frac{a^{(#1)}_{#2}}{a^{(#3)}_{#4}}}}
\newcommand{\xb}[3]{\ensuremath{x_{#1}^{(#2,#3)}}}
\newcommand{\yb}[3]{\ensuremath{y_{#1}^{(#2,#3)}}}

\def\identity{{\rlap{1} \hskip 1.6pt \hbox{1}}}
\def\iden{\identity}

\def\IB{\mathbb{B}}
\def\IC{\mathbb{C}}
\def\ID{\mathbb{D}}
\def\IH{\mathbb{H}}
\def\IM{\mathbb{M}}
\def\IN{\mathbb{N}}
\def\IP{\mathbb{P}}
\def\IR{\mathbb{R}}
\def\IZ{\mathbb{Z}}

\def\CA{{\cal A}}
\def\CB{{\cal B}}
\def\CC{{\cal C}}
\def\CD{{\cal D}}
\def\CE{{\cal E}}
\def\CF{{\cal F}}
\def\CG{{\cal G}}
\def\CH{{\cal H}}
\def\CI{{\cal I}}
\def\CJ{{\cal J}}
\def\CK{{\cal K}}
\def\CL{{\cal L}}
\def\CM{{\cal M}}
\def\CN{{\cal N}}
\def\CO{{\cal O}}
\def\CP{{\cal P}}
\def\CQ{{\cal Q}}
\def\CR{{\cal R}}
\def\CS{{\cal S}}
\def\CT{{\cal T}}
\def\CU{{\cal U}}
\def\CV{{\cal V}}
\def\CW{{\cal W}}
\def\CX{{\cal X}}
\def\CY{{\cal Y}}
\def\CZ{{\cal Z}}

\def\a{\alpha}
\def\b{\beta}
\def\g{\gamma}
\def\d{\delta}
\def\e{\epsilon}
\def\ve{\varepsilon}
\def\z{\zeta}
\def\th{\theta}
\def\vth{\vartheta}
\def\i{\iota}
\def\k{\kappa}
\def\l{\lambda}
\def\m{\mu}
\def\n{\nu}
\def\vp{\varpi}
\def\r{\rho}
\def\vr{\varrho}
\def\s{\sigma}
\def\vs{\varsigma}
\def\t{\tau}
\def\u{\upsilon}
\def\vph{\varphi}
\def\ch{\chi}
\def\w{\omega}
\def\G{\Gamma}
\def\D{\Delta}
\def\Th{\Theta}
\def\L{\Lambda}
\def\S{\Sigma}
\def\Y{\Upsilon}
\def\O{\Omega}

\newcommand{\chris}[2]{\Gamma^{#1}_{\,\,\,\,#2}}

\def\half{\frac{1}{2}}
\def\imp{\Longrightarrow}
\def\goto{\rightarrow}
\def\para{\parallel}
\def\vev#1{\langle #1 \rangle}
\def\del{\nabla}
\def\p{\partial}
\newcommand{\bra}[1]{\langle{#1}|}
\newcommand{\ket}[1]{|{#1}\rangle}

\newcommand{\suup}{1}
\newcommand{\sudown}{2}

\def\Tr{{\rm Tr}}
\def\tr{{\rm Tr}}
\def\det{{\rm det}}
\def\PE{{\rm PE}}

\def\fa{\mathfrak{a}}
\def\fb{\mathfrak{b}}
\def\fc{\mathfrak{c}}
\def\fp{\mathfrak{p}}
\def\fq{\mathfrak{q}}
\def\ft{\mathfrak{t}}
\def\fG{\mathfrak{G}}
\def\PE{\textrm{PE}}

\def\Xint#1{\mathchoice
{\XXint\displaystyle\textstyle{#1}}%
{\XXint\textstyle\scriptstyle{#1}}%
{\XXint\scriptstyle\scriptscriptstyle{#1}}%
{\XXint\scriptscriptstyle\scriptscriptstyle{#1}}%
\!\int}
\def\XXint#1#2#3{{\setbox0=\hbox{$#1{#2#3}{\int}$}
\vcenter{\hbox{$#2#3$}}\kern-.5\wd0}}
\def\ddashint{\Xint=}
\def\dashint{\Xint-}

\title{D-type fiber-base duality}

\begin{document}
\author[\ast]{Babak Haghighat,}
\author[\dag]{Joonho Kim,}
\author[\ast]{Wenbin Yan,}
\author[\ddagger]{Shing-Tung Yau}
\affiliation[\ast]{Yau Mathematical Sciences Center, Tsinghua University, Beijing, 100084, China}
\affiliation[\dag]{School of Physics, Korea Institute for Advanced Study, Seoul 02455, Korea}
\affiliation[\ddagger]{Department of Mathematics, Harvard University, Cambridge, 02138, USA}

\abstract{M5 branes probing D-type singularities give rise to 6d (1,0) SCFTs with $SO \times SO$ flavor symmetry known as D-type conformal matter theories. Gauging the diagonal $SO$-flavor symmetry leads to a little string theory with an intrinsic scale which can be engineered in F-theory by compactifying on a doubly-elliptic Calabi-Yau manifold. We derive Seiberg-Witten curves for these little string theories which can be interpreted as mirror curves for the corresponding Calabi-Yau manifolds. Under fiber-base duality these models are mapped to D-type quiver gauge theories and we check that their Seiberg-Witten curves match. By taking decompactification limits, we construct the curves for the related 6d SCFTs and connect to known results in the literature by further taking 5d and 4d limits.}

\begin{flushright}
KIAS-P18071
\end{flushright}
	
\maketitle

\section{Introduction}

The classification of 6d $\mathcal{N}=(1,0)$ SCFTs in F-theory through elliptic Calabi-Yau manifolds \cite{Heckman:2013pva,Heckman:2015bfa} naturally leads to the question of classification of 4d vacua obtained from these theories in a dimensional reduction. The most straightforward direction to proceed  is to construct the Seiberg-Witten curves of the resulting four-dimensional $\mathcal{N}=2$ theories upon compactification on a two-torus. This approach has been pursued in \cite{DelZotto:2015rca,Ohmori:2015pua,Ohmori:2015pia} using different methods. The method used in \cite{DelZotto:2015rca} is the orbifold Landau-Ginzburg technique, while the strategy of \cite{Ohmori:2015pua,Ohmori:2015pia} has been the connection to $(2,0)$ compactifications on Gaiotto curves. Both methods have limited scopes while shedding light on different aspects of the compactification. Orbifold Landau-Ginzburg models can be applied to any F-theory compactification which admits an orbifold description as a discrete quotient of $T^2 \times \mathbb{C} \times \mathbb{C}$. Thus the method has been successfully applied to the non-Higgsable classes with one tensor multiplet and to various conformal matter theories. The procedure involves constructing the mirror geometry of the orbifold $(T^2\times \mathbb{C} \times \mathbb{C})/\Gamma_\mathrm{G}$ which leads to the SW-curve of an intrinsically four-dimensional theory and then taking a limit in moduli space to reach a CFT point. In constructing the resulting theories, however, a certain limit of the Calabi-Yau geometry has been taken which from the 6d SCFT point of view involves shrinking the radii of compactification from 6d to 4d. On the other hand, the approach of \cite{Ohmori:2015pua,Ohmori:2015pia} has been to identify a quiver description for the reduced theory and subsequently use the technique of associating a Gaiotto curve to such a theory from which the SW-curve can be read off. 

In the present paper we will be taking yet another direction to construct the Seiberg-Witten curve. Our approach is based on the recent progress in computing 6d SCFT BPS partition functions by identifying the 2d theories on the worldvolume of strings which appear on their tensor branch \cite{Haghighat:2013gba,Haghighat:2013tka,Haghighat:2014pva,Kim:2014dza,Haghighat:2014vxa,Gadde:2015tra,Kim:2015fxa,Kim:2016foj,DelZotto:2016pvm,Kim:2018gjo, Kim:2018gak,DelZotto:2018tcj}. As advocated in \cite{Haghighat:2017vch}, we propose to take the thermodynamic limit \cite{Nekrasov:2003rj} of 6d partition functions in order to obtain the SW-curve as the spectral curve of the resulting matrix model. Given the recent advances in computing these partition functions, it seems natural to pursue this path as more non-trivial SCFTs, which do not admit an orbifold description and do not connect to other $(2,0)$ theories upon compactification, move within reach. Moreover, in this approach we can keep all radii in the game finite and thus obtain an expression for the true 6d curve which still depends non-trivially on the 6d to 5d and 5d to 4d compactification radii. Given such a more general curve, it is expected that its singular loci in moduli space classify the corresponding 4d SCFTs which can be reached. 

We will be looking at 6d $(1,0)$ SCFTs which arise from M5 branes probing $D$-type singularities and the SW-curves they give rise to. As it turns out, in order to derive the equations for the curve, it is useful to ``uplift" these theories to little string theories by compactifying the chain of $\mathbb{P}^1$'s in the base of the Calabi-Yau geometries to an elliptic curve, thus making the Calabi-Yau doubly elliptic. The little string theory obtained this way admits fiber-base duality which is essentially T-duality transforming the system of M5 branes to D5 branes probing a $D$-type singularity. The SW-curve for this T-dual picture was obtained in \cite{Haghighat:2017vch} by generalizing the construction of \cite{Nekrasov:2012xe}. We analyze the special case of one M5 brane and one D5 brane in detail and show that the two SW-curves are indeed identical. We can then take two different SCFT limits by decompactifying the base either by sending the volume of a $-1$-curve to infinity or that of a $-4$-curve. This decompactification limit not only fixes the form of the SCFT SW-curve but also gives further information about the structure of the SW-curve of the little string theory which in turn gives further consistency checks for our ansatz. We show that so obtained 6d curves are indeed the most general curves as they correctly reproduce the known 5d and 4d curves upon sending the compactification radii to zero. 

One remarkable aspect of the little string curves is that they can be given interpretations of spectral curves corresponding to moduli spaces of instantons on a complex two-dimensional torus (also known as an abelian surface) on the one hand and instantons on a particular K3 surface on the other hand. In fact, the moduli space of quantum vacua of the two little string theories can be identified with the two corresponding instanton moduli spaces. In the case of M5 branes and D5 branes probing an $A$-type singularity this correspondence is well-known to mathematicians and is the one between moduli spaces of $N$ $SU(r)$ instantons on $T^4$ on the one side and that of $r$ $SU(N)$ instantons on the other \cite{Jardim:2000cr}. But whereas in the $A$-type case the abelian surface is a general one, in the $D$ and $E$ type cases it is restricted to be a product of two elliptic curves\footnote{What makes the $D$ and $E$ cases more complicated is the equation defining the curve inside the abelian surface.}, the reason being that the $A$-type singularity has a further $U(1)$ isometry corresponding to mass deformation which the other two cases do not admit. One way to view the results of this paper is a generalization of this correspondence to $SO(8)$-instantons on $T^4$ and $SU(2)$ instantons on $K3 \equiv T^4/\mathbb{Z}_2$.

The organization of the paper is as follows. In Section 2 we review the brane construction for the two little string theories in question. In Section 3 we give a precise account on the thermodynamic limit of the 6d partition function and derive saddle-point equations in this limit which define the spectral curve. We then proceed in Section 4 to derive concrete expressions for the spectral curve. We observe the invariance under fiber-base duality by comparing to the dual SW-curve obtained by taking the thermodynamic limit of another little string partition function. Finally, we take the SCFT limits as well as 5d and 4d limits of the obtained curve and compare with existing results. We end the paper with a discussion giving an outlook on open problems and directions to proceed.

\section{Brane construction}

In this section we want to review the brane construction for the little string theories (LSTs) of interest and their duality frames. This will allow us to describe their quantum moduli spaces of vacua in a coherent formalism and look at various limiting behaviors obtained by successively sending radii involved to zero. These limits will correspond to the 6d SCFT, 5d and 4d limits of the theory. The LSTs we will be interested in arise on the one hand in Type IIB string theory from D5-branes probing ADE singularities and on the other hand from M5 branes probing ADE singularities. The two constructions, denoted by $\mathcal{T}^B$ and $\mathcal{T}^A$ respectively, are related through fiber-base duality of doubly elliptic Calabi-Yau threefolds \cite{Bhardwaj:2015oru} which can equally well be interpreted as T-duality in Type II string theory \cite{Ohmori:2015pia}.

\subsection*{D5 branes probing ADE singularities}

\begin{figure}[h]
  \centering
	\includegraphics[width=\textwidth]{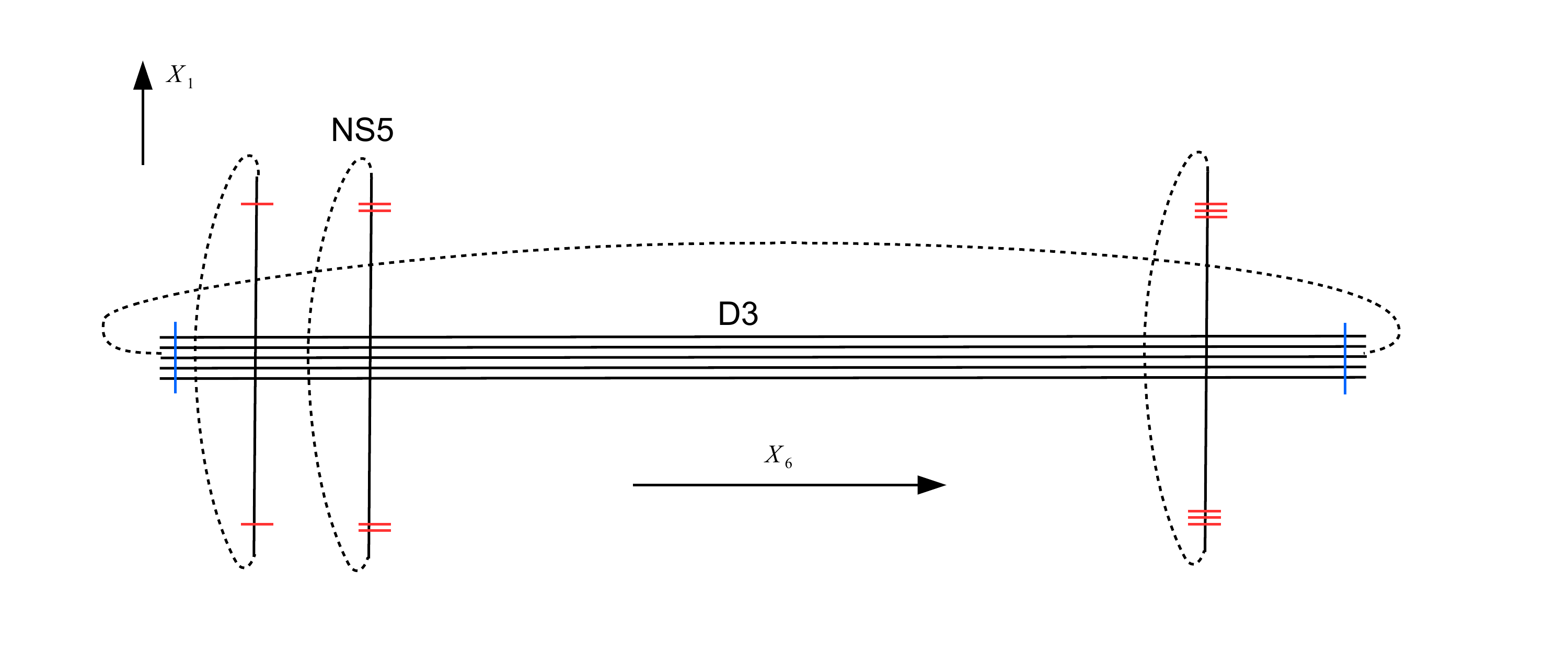}
  \caption{Brane setup for a 3d $\mathcal{N}=4$ theory capturing the Coulomb branch of our original $\mathcal{T}^B_{A_r}$ theory.}
  \label{fig:Abranes}
\end{figure}

This case corresponds to the $\mathcal{T}^B$ theories \cite{Blum:1997mm}. Let us focus for simplicity on A-type singularities and then successively generalize from results draws from this case. The brane configuration in this case is shown in the following table:
\begin{center}
\begin{tabular}{c|cccccccccc}
& $S^1$ & $S^1$ & \multicolumn{4}{c}{$\mathbb{R}^4_\parallel$} &\multicolumn{4}{c}{$TN_{r+1}$}\\
&  $ X^0 $& $ X^1 $& $ X^2 $& $ X^3 $& $ X^4 $& $ X^5 $& $ X^6 $& $ X^7 $& $ X^8 $ & $X^9$\\
\hline
$N$ D5 & $\times$ & $\times$ & $\times$ &$\times$ &$\times$ & $\times$ & -- & -- & --& --
\end{tabular}
\end{center}
We want to deduce the moduli space of vacua corresponding to the Coulomb branch of the 4d $\mathcal{N}=2$ compactification. To this end, we perform T-duality along the Taub-NUT circle and arrive at the type IIA brane configuration:
\begin{center}
\begin{tabular}{c|cccccccccc}
& $S^1$ & $S^1$ & \multicolumn{4}{c}{$\mathbb{R}^4_\parallel$} &$S^1$ & \multicolumn{3}{c}{$\mathbb{R}^3$}\\
&  $ X^0 $& $ X^1 $& $ X^2 $& $ X^3 $& $ X^4 $& $ X^5 $& $ X^6 $& $ X^7 $& $ X^8 $ & $X^9$\\
\hline
$r+1$ NS5 & $\times$ & $\times$ & $\times$ &$\times$ &$\times$ & $\times$ & $\{y_i\}$ & -- & --& --\\
$N$ D6 & $\times$ & $\times$ & $\times$ & $\times$ & $\times$ & $\times$ & $\times$ & -- & -- & --
\end{tabular}
\end{center}
We next perform T-duality along the $X^1$ circle and arrive at the following type IIB brane setup:
\begin{center}
\begin{tabular}{c|cccccccccc}
& $S^1$ & $\widetilde{S}^1$ & \multicolumn{4}{c}{$\mathbb{R}^4_\parallel$} &$S^1$ & \multicolumn{3}{c}{$\mathbb{R}^3$}\\
&  $ X^0 $& $ X^1 $& $ X^2 $& $ X^3 $& $ X^4 $& $ X^5 $& $ X^6 $& $ X^7 $& $ X^8 $ & $X^9$\\
\hline
$r+1$ NS5 & $\times$ & $\times$ & $\times$ &$\times$ &$\times$ & $\times$ & $\{y_i\}$ & -- & --& --\\
$N$ D5 & $\times$ & -- & $\times$ & $\times$ & $\times$ & $\times$ & $\times$ & -- & -- & --
\end{tabular}
\end{center}
In order to arrive at a description of the Coulomb branch, we further compactify the direction $X^2$. The moduli space is then the one of the resulting 3d $\mathcal{N}=4$ theory which admits a Hitchin system description. To deduce it, we perform two T-dualities along $X^0$ and $X^2$ and arrive at the picture shown in Figure \ref{fig:Abranes}. 
The D5 branes have now become D3 branes which form impurities in the gauge theory living on the NS5 branes. In the current case, due to the $A_r$ singularity we started with, this gauge group is $G_{A_{r}} = SU(r+1)$. There are now several limits we can consider:\footnote{The radii $R_0$ and $R_1$ are the ones of the original configuration corresponding to the 6d SCFT shown in the first table before applying the T-dualities. The T-dual variables have to be sent to $\infty$ in this limit.} \\

\begin{tabular}{lll}
1) & $R_0, R_1 \rightarrow 0$ and $R_6 \rightarrow \infty$ & D3-branes are periodic monopoles on $\mathbb{R}^2 \times S^1$ \\
2) & $R_2 \rightarrow 0$ and $R_6 \rightarrow \infty$ & D3-branes are doubly-periodic monopoles on $\mathbb{R} \times T^2$ \\
3) & $R_6 \rightarrow \infty$ & D3-branes are triply-periodic monopoles on $T^3$\\
4) & all radii are finite & Instantons on $T^4$
\end{tabular}\\

\noindent 
We refer to \cite{Cherkis:2014vfa} for more details on the moduli spaces of periodic monopoles. To see how the fourth case comes about, note that D3-branes are S-duality invariant. Thus performing S-duality and then successively T-duality along $X^6$, we arrive at D2-branes as instantons in D6-branes wrapping a four-torus composed of the periodic directions $X^0, X^1, X^2, X^6$. In the original D5 setup the above limits correspond to:\\

\begin{tabular}{ll}
1) & 4d quiver gauge theory on $\mathbb{R}^3 \times S^1$ \\
2) & 5d quiver gauge theory on $\mathbb{R}^3 \times T^2 $ \\
3) & Tensor branch of 6d SCFT on $\mathbb{R}^3 \times T^3$\\
4) & 6d LST on $\mathbb{R}^3 \times T^3$
\end{tabular}\\

\noindent
It is the last case which is of interest for us in this paper. Let us conclude by noting that the generalization of this construction to the $D$ and $E$ types amounts to identifying the moduli spaces of the corresponding LSTs with those of $SO(8+2n)$ and $E_6$, $E_7$, $E_8$ instantons on $T^4$ \cite{Intriligator:1999cn}.

\subsection*{M5 branes probing ADE singularities}

\begin{figure}[h]
  \centering
	\includegraphics[width=\textwidth]{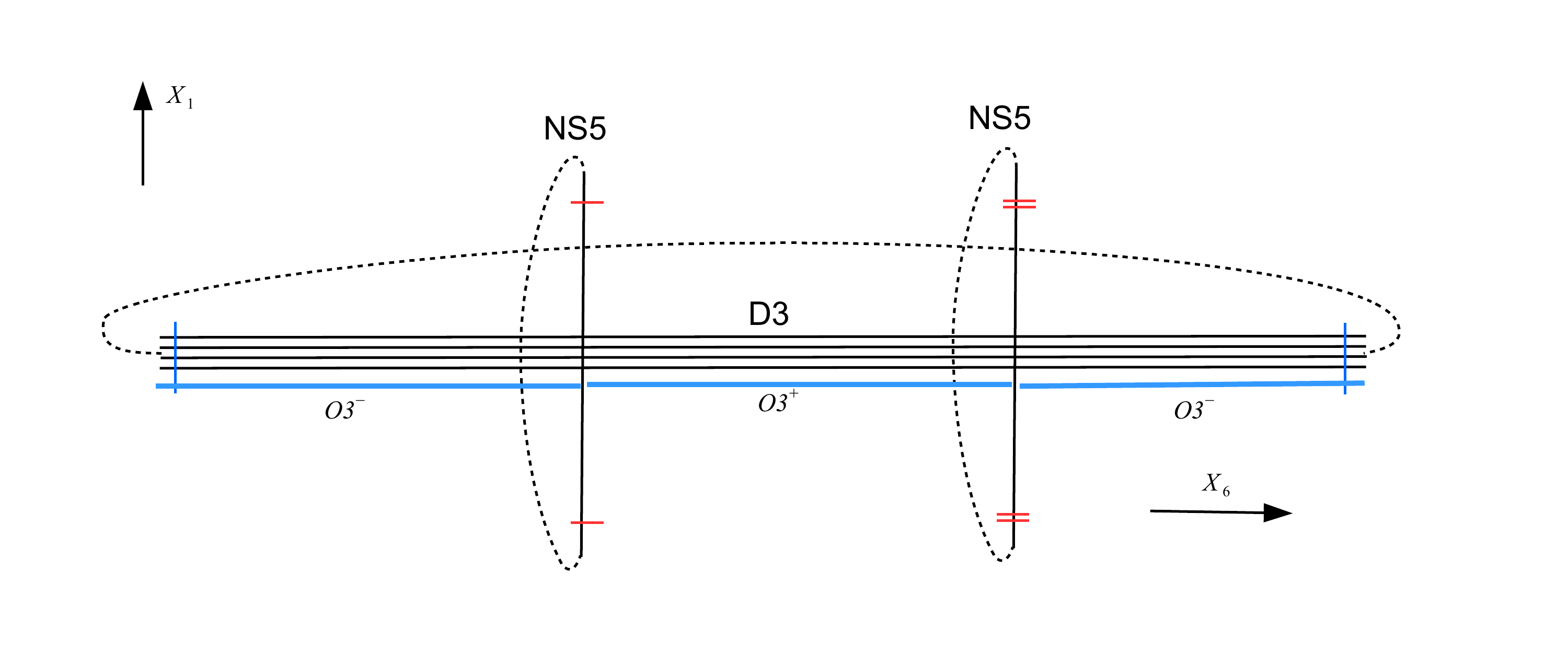}
  \caption{Brane setup for a 3d $\mathcal{N}=4$ theory capturing the Coulomb branch of our original $\mathcal{T}^A_{D_4}$ theory.}
  \label{fig:Dbranes}
\end{figure}

We now turn our attention to the LSTs arising from M5 branes probing ADE singularities, denoted $\mathcal{T}^A$ \cite{DelZotto:2014hpa,Bhardwaj:2015oru}. The $A$-type case is already covered in the construction above which can be identified as its T-dual. The central example of this paper will be the $D$-type singularity on which we want to focus in the following. Let us first focus on the case of a single M5 brane probing a $D_4$ singularity:
\begin{center}
\begin{tabular}{c|ccccccccccc}
& $S^1$ & $S^1$ & \multicolumn{4}{c}{$\mathbb{R}^4_\parallel$} & $S^1$ & \multicolumn{4}{c}{$\mathbb{C}^2/\Gamma_{D_4}$}\\
&  $ X^0 $& $ X^1 $& $ X^2 $& $ X^3 $& $ X^4 $& $ X^5 $& $ X^6 $& $ X^7 $& $ X^8 $ & $X^9$ & $X^{10}$\\
\hline
M5 & $\times$ & $\times$ & $\times$ &$\times$ &$\times$ & $\times$ & -- & -- & --& -- & --
\end{tabular}
\end{center}
In this case it is known that the M5 brane fractionates into $2$ half M5 branes along the $X^6$ circle. By reducing the ALF circle, they become IIA NS5-branes between which we have D6-branes with $O6^+$ and $O6^-$ on different sides of the NS5 branes \cite{Evans:1997hk,Hanany:1997gh,Brunner:1997gf}, thus giving the following setup:
\begin{center}
\begin{tabular}{c|ccccccccccc}
& $S^1$ & $S^1$ & \multicolumn{4}{c}{$\mathbb{R}^4_\parallel$} & $S^1$ & \multicolumn{3}{c}{$\mathbb{R}^3_{\perp}$}\\
&  $ X^0 $& $ X^1 $& $ X^2 $& $ X^3 $& $ X^4 $& $ X^5 $& $ X^6 $& $ X^7 $& $ X^8 $ & $X^9$ \\
\hline
NS5 & $\times$ & $\times$ & $\times$ &$\times$ &$\times$ & $\times$ & $y_1$ & -- & --& -- \\
NS5 & $\times$ & $\times$ & $\times$ &$\times$ &$\times$ & $\times$ & $y_2$ & -- & --& -- \\
$O6^+$ & $\times$ & $\times$ & $\times$ &$\times$ &$\times$ & $\times$ & $[y_1,y_2)$ & -- & --& -- \\
$O6^-$ & $\times$ & $\times$ & $\times$ &$\times$ &$\times$ & $\times$ & $[y_2,y_1)$ & -- & --& -- \\
$4$ D6 & $\times$ & $\times$ & $\times$ &$\times$ &$\times$ & $\times$ & $\times$ & -- & --& -- 
\end{tabular}
\end{center}
Next, we compactify the $X^2$ direction and perform 3 T-dualities along $X^0$, $X^1$ and $X^2$ which are common to the NS5 branes and D6 branes. We end up with D3 branes on top of $O3^{\pm}$ planes suspended between NS5 branes as shown in Figure \ref{fig:Dbranes}.
 Now we perform S-duality of type IIB and then another T-duality along the $X^6$ direction and end up with D2 and O2 planes in D6 branes wrapping the periodic directions $X^0$, $X^1$, $X^2$ and $X^6$. In fact, the orientifolds fractionate and we end up with $16$ O2 planes located at the  $16 = 4 \times 4$ fixed points of $T^4 \cong T^2 \times T^2$ under the $\mathbb{Z}_2$ action $X^i \mapsto - X^i$. Thus we arrive at $2$ D6 branes wrapping the K3 surface $T^4/\mathbb{Z}_2$ together with $4$ D2 instantons of the corresponding $SU(2)$ gauge theory. This picture can easily be generalized to the case of $N$ M5 branes probing a $D_{4+n}$ singularity. The corresponding Coulomb branch moduli space is then the one of $4+n$ $SU(2N)$ instanton on $T^4/\mathbb{Z}_2$.
 
 \subsection*{Spectral curves}
 
Let us first focus on the case of $N$ D5 branes probing an affine $A_r$ singularity. The four-torus appearing in the above constructions can be viewed as an elliptic surface by identifying one torus as its fiber and the other as its base. Thus we will be writing $T^4 \cong \mathbb{T}_{\tau_f} \times \mathbb{T}_{\tau_b}$, where $\mathbb{T}_{\tau} = \mathbb{C}/(\mathbb{Z}\oplus \tau \mathbb{Z})$. Now the moduli space of $N$ $SU(r+1)$ instantons on $T^4$ can be identified with the moduli space of the \textit{instanton spectral curve} with respect to the projection $X \stackrel{\pi}{\rightarrow} \mathbb{T}_{\tau_b}$. Such a curve is a branched $r$-fold covering map where $r$ is the rank of the gauge group $G$. It is given in terms of the zero set of the determinant of a $G$-bundle over the Jacobian of the fiber $\widehat{\mathbb{T}}_{\tau_f} = H^1(\mathbb{T}_{\tau_f},\mathbb{R})/H^1(\mathbb{T}_{\tau_f},\mathbb{Z})$ (which is isomorphic to $\mathbb{T}_{\tau_f}$ itself) \cite{Friedman:1997yq}. Thus we see that the spectral curve $S$ has a realization as a holomorphic curve inside $\widehat{\mathbb{T}}_{\tau_f} \times \mathbb{T}_{\tau_b}$. On the other hand, the moduli space of the T-dual description in terms of $N$ M5 branes probing an affine $A_r$ singularity is the one of $r$ instantons of $SU(N)$ gauge theory on the dual torus $\widehat{T}^4 \cong  \widehat{\mathbb{T}}_{\tau_f} \times \widehat{\mathbb{T}}_{\tau_b}$. In this case $\widehat{\mathbb{T}}_{\tau_b}$ is the fiber and the spectral curve is a holomorphic curve inside $\widehat{\mathbb{T}}_{\tau_f} \times \widehat{\widehat{\mathbb{T}}}_{\tau_b} = \widehat{\mathbb{T}}_{\tau_f} \times \mathbb{T}_{\tau_b}$. In fact, the two spectral curves are identical and one can show that the two descriptions are related through the so called \textit{Fourier-Mukai} transform \cite{Jardim:2000cr}. Hence there is a bijective correspondence between the two instanton moduli spaces: $\mathcal{M}_{T^4}(r+1,N) \leftrightarrow \mathcal{M}_{\widehat{T}^4}(N,r+1)$. As shown in \cite{Haghighat:2017vch}, such a spectral curve can be interpreted as the mirror curve of the elliptic Calabi-Yau manifolds engineering the corresponding 6d LSTs in F-theory. In the $A$-type case the mirror curve is given in terms of a linear combination of genus $2$ Riemann theta functions and the Fourier-Mukai transform in that context corresponds to the element
\begin{equation}
 \left(\begin{array}{cc} 0 & \mathbf{1} \\ - \mathbf{1} & 0 \end{array}\right) \in Sp(4,\mathbb{Z}),
\end{equation}
swapping the 2-tori inside the four-torus.

Let us next come to the $D$-type singularity. The case of a D5-brane probing a $D_4$ singularity corresponds to the moduli space of an $SO(8)$ instanton on $T^4$. The spectral curve is living in $\widehat{\mathbb{T}}_{\tau_f} \times \mathbb{T}_{\tau_b}$ and restricted to the fiber, it is given as the zero set of the determinant of an $SO(8)$-bundle over $\widehat{\mathbb{T}}_{\tau_f}$ \cite{Haghighat:2017vch}. On the other hand, in the dual picture we have a non-trivial fibration of $\mathbb{T}_{\tau_b}$ over $\widehat{\mathbb{T}}_{\tau_f}/\mathbb{Z}_2$ such that the fiber degenerates over four fixed-points of the $\mathbb{Z}_2$ action in the base. As we will see in the following sections, when restricted to the fiber, the spectral curve turns out to be the zero set of the determinant of an $SL(2)$ bundle. Moreover, we will see how the $SO(8)$ bundle of the previous picture arises from the four instantons and their mirror images under the $\mathbb{Z}_2$ action. The procedure we will be using to arrive at the spectral curve is the thermodynamic limit of 
\cite{Nekrasov:2003rj}, to which we turn now.

\section{The thermodynamic limit}

\begin{figure}[h!]
\centering
\includegraphics[scale=0.7]{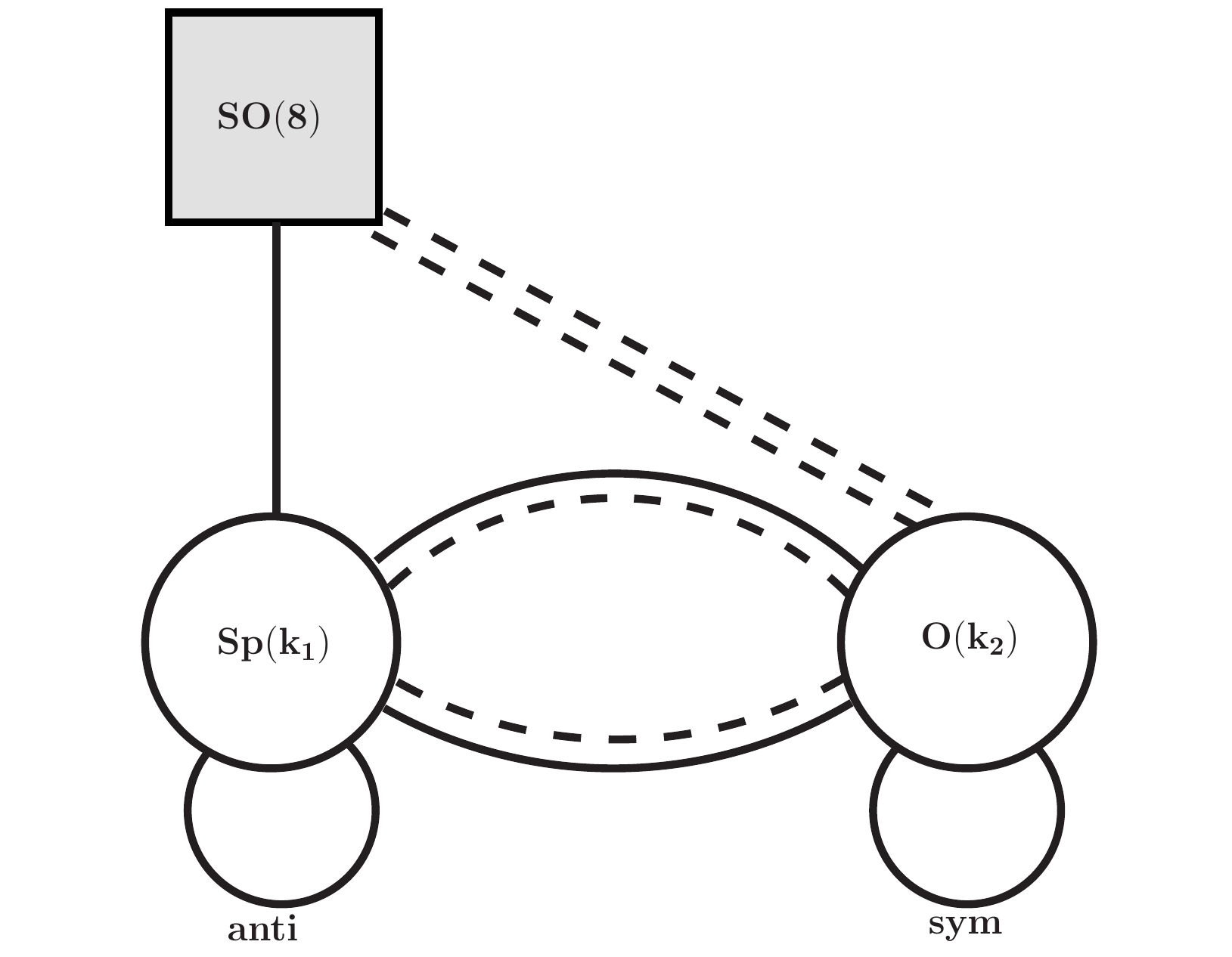}
\caption{\label{fig:quiverD4N1}Quiver for one instanton on $D_4$. }
\end{figure}

The thermodynamic limit, as developed in \cite{Nekrasov:2003rj} and further applied in \cite{Nekrasov:2012xe} to the case of quiver gauge theories, provides a straightforward derivation of Seiberg-Witten curves once there is  sufficient information for the instanton sector of a gauge theory. The procedure involves writing the Nekrasov partition function, which is equivalent to the topological string partition function of the Calabi-Yau manifolds engineering the gauge theory, as a sum over its instanton sectors:
\begin{equation}
	Z^{top} = e^{\sum_g \hbar^{2g-2} F_g} = Z^{pert} \left(1 + \sum_{\{k_i\}} q_1^{k_1} q_2^{k_2} \ldots q_r^{k_r} Z_{\{k_i\}}\right),
\end{equation}
where $\hbar$ is the topological string coupling constant and $q_i = e^{2\pi i \tau_i}$ with $\tau_i$ being the complexified gauge coupling of the $i$th gauge node. The $Z_{\{k_i\}}$ can be computed through supersymmetric localization on instanton moduli spaces and can themselves be written in terms of discrete sums. In the limit where the topological string coupling constant goes to zero, $\hbar \rightarrow 0$, the above sum becomes a path integral of the following form
\begin{equation}
	Z^{top} \sim \int \prod_i \mathcal{D}\varrho_i e^{\frac{\mathcal{F}_0(\vec{\tau},\vec{\varrho})}{\hbar} + \mathcal{O}(1)},
\end{equation}
where the $\varrho_i$ can be viewed as eigenvalue densities of the $i$th instanton gauge group (not to be confused with the bulk gauge group). This leads to a matrix model from which we can extract the spectral curve. The instanton sector of the little string theory $\mathcal{T}^A_{D_4,1}$ (i.e. one M5 brane probing a $D_4$ singularity) compactified on $\mathbb{R}^4 \times T^2$ is captured by the 2d quiver gauge theory shown in Figure \ref{fig:quiverD4N1}.
It is composed of an E-string node \cite{Kim:2014dza} and an $\mathcal{O}(-4)$ node \cite{Haghighat:2014vxa} combined into a circular $-1$,$-4$ chain \cite{Kim:2017xan}. $k_1$ and $k_2$ denote the winding numbers of two fractional little strings corresponding to the $-4$ and $-1$ curves in the F-theory construction. The topological string partition function in this case then boils down to computing the following infinite sum
\begin{equation}
Z^{top}=Z^{pert}\left(1+\sum_{k_1, k_2\geq0}q_1^{k_1}q_0^{k_2}\sum_d Z^{Sp(k_1) \times O(k_2),d}\right),
\end{equation}
where $\sum_d Z^{Sp(k_1) \times O(k_2),d}$ denotes the elliptic genus of a string chain composed of $k_1$ $SO(8)$ instanton strings and $k_2$ E-strings.
We refer to \cite{Kim:2017xan} for its precise definition.
Notice that care has been taken of the fact that it has discrete gauge moduli labeled by $d$ in the superscript of the elliptic genus. In the following we want to find an integral representation for the elliptic genera, whose integrand is known as the $1$-loop determinant. 

\subsection{The 1-loop determinants}
The field content of the quiver depicted in Figure \ref{fig:quiverD4N1} is shown in Table \ref{table:D4N1fields}. 
\begin{table}[htbp]
\centering
\begin{tabular}{c|c|c}
Type & Fields & Representation \\
\hline
vector & $(A_\mu,\lambda^{\dot{\alpha}A}_-)$ & $\mathbf{adj}$ of $Sp(k_1)$ \\
hyper & $(a_{\alpha\dot{\alpha}},\psi^{\alpha A}_+)$ &$\mathbf{anti}$ of $Sp(k_1)$ \\
hyper & $(q_{\dot{\alpha}},\psi^{A}_+)$ & $\mathbf{bif}$ of $Sp(k_1)\times SO(8)$\\
\hline
vector & $(A_\mu,\lambda^{\dot{\alpha}A}_-)$ & $\mathbf{adj}$ of $O(k_2)$ \\
hyper & $(a_{\alpha\dot{\alpha}},\psi^{\alpha A}_+)$ & $\mathbf{sym}$ of $O(k_2)$ \\
Fermi & $(\chi_1),(\chi_2)$ & $\mathbf{bif}$ of $O(k_2)\times SO(8)$\\
\hline
twisted hyper & $(\phi_{\dot{\alpha}},\mu^A_+)_1$ & $\mathbf{bif}$ of $Sp(k_1)\times O(k_2)$\\
Fermi & $(\mu^\alpha_-)_1$ & $\mathbf{bif}$ of $Sp(k_1)\times O(k_2)$\\
\hline
twisted hyper & $(\phi_{\dot{\alpha}},\mu^A_+)_2$ & $\mathbf{bif}$ of $O(k_2)\times Sp(k_1)$\\
Fermi & $(\mu^\alpha_-)_2$ & $\mathbf{bif}$ of $O(k_2)\times O(k_1)$
\end{tabular}
\caption{\label{table:D4N1fields}The field contents of $D_4$.}
\end{table}
The  corresponding theory is a 2d $\mathcal{N} = (0,4)$ gauge theory and the 1-loop determinants over the superfields are,
\begin{equation}
\label{eq:1loop}
\begin{split}
Z_{\mathrm{vector}}&=\prod_{i=1}^r\,2\pi\eta^2\,\theta_1(2\epsilon_+)d\phi_i
\prod_{\rho\in\mathrm{root}}\frac{\theta_1(\rho\cdot\phi)\theta_1(\rho\cdot\phi+2\epsilon_+)}{\eta^2},\\
Z_{\mathrm{Fermi}}&=\prod_{\rho\in\mathrm{rep_g}}
\prod_{\kappa\in\mathrm{rep_f}}\frac{\theta_1(\rho\cdot\phi+\kappa\cdot z)}{\eta},\\
Z_{\mathrm{hyper}}&=\prod_{\rho\in\mathrm{rep_g}}
\prod_{\kappa\in\mathrm{rep_f}}\frac{\eta}{\theta_1(\epsilon_++\rho\cdot\phi+\kappa\cdot z)},\\
Z_{\mathrm{twisted hyper}}&=\prod_{\rho\in\mathrm{rep_g}}
\prod_{\kappa\in\mathrm{rep_f}}\frac{\eta}{\theta_1(-\epsilon_++\rho\cdot\phi+\kappa\cdot z)}.
\end{split}
\end{equation}
$\phi$ denotes the gauge holonomy eigenvalue, $\rho$ is the eigenvalue of the Cartan generator of the gauge symmetry in the representation $\mathrm{rep_g}$. $\kappa$ collectively denotes the eigenvalues for the Cartan generators of all global symmetry including $SU(2)_{L}$, i.e., anti-self-dual rotation of the  $\mathbb{R}^4$.  We refer to the appendix for our conventions of the Jacobi theta function and the eta function as well as other modular forms appearing in this paper.\footnote{For reasons of clarity in the presentation, we will omit the dependence on the modular parameter in all theta-functions appearing in this section.}
\subsubsection*{The $Sp(k_1)$ node}
The contribution to the 1-loop determinant from the pure $Sp(k_1)$ part is \begin{equation}
Z^{Sp(k_1)}=Z^{Sp(k_1)}_{\mathrm{vector}}Z^{Sp(k_1)}_{\mathrm{hyper},\mathbf{anti}}Z^{Sp(k_1)}_{\mathrm{hyper}},
\end{equation}
where
\begin{align}
\label{eq:sp-1-loop}
Z^{Sp(k_1)}_{\mathrm{vector}}
&=\prod_{i=1}^{k_1}2\pi\eta^2\,\theta_1(2\epsilon_+)d\phi_i
\frac{\theta_1(\pm2\phi_i)\theta_1(\epsilon_+\pm2\phi_i)}{\eta^4}
\prod_{1\leq i<j\leq k_1}\frac{\theta_1(\pm\phi_i\pm\phi_j)\theta(2\epsilon_+\pm\phi_i\pm\phi_j)}{\eta^8},\nonumber\\
Z^{Sp(k_1)}_{\mathrm{hyper},\mathbf{anti}}
&=\prod_{1\leq i<j\leq k_1}\frac{\eta^8}{\theta_1(\epsilon_++\epsilon_-\pm\phi_i\pm\phi_j)\theta(\epsilon_+-\epsilon_-\pm\phi_i\pm\phi_j)},\nonumber\\
Z^{Sp(k_1)}_{\mathrm{hyper}}
&=\prod_{i=1}^{k_1}\prod_{j=1}^{4}\frac{\eta^4}{\theta_1(\epsilon_+\pm\phi_i\pm m_j)},
\end{align}
and $\e_+=\half(\e_1+\e_2)$ and $\e_-=\half(\e_1-\e_2)$.
To derive \eqref{eq:sp-1-loop} from \eqref{eq:1loop}, we used the fact that the gauge holonomy $\phi$ is a $2k_1$ by $2k_1$ matrix with the following eigenvalues:
\begin{equation}
\phi = \mathrm{diag}(\pm\phi_1,\pm\phi_2,\cdots,\pm\phi_{k_1}).
\end{equation}
The $2k_1$-dimensional $Sp(k_1)$ charge vectors are given by
\begin{align}
\rho &=(0,\cdots,0,1,0,\cdots,0) && \text{for the defining representation},\\
\rho&=(0,\cdots,0,2,0,\cdots,0)\ \ \mathrm{or} \nn\\
&\ \ \ \ (0,\cdots,0,1,0,\cdots,0,1,0,\cdots,0) \nn
&& \text{for the adjoint representation},\\
\rho&=(0,\cdots,0,1,0,\cdots,0,1,0,\cdots,0)  && \text{for the antisymmetric representation}, \nn
\end{align}
from which one can compute $(\rho \cdot \phi)$ and then obtain the expression \eqref{eq:sp-1-loop}. 

\subsubsection*{The $O(k_2)$ node}
The contribution from the pure $O(k_2)$ (assuming $k_2$ even) part is
\begin{equation}
Z^{O(k_2)}=Z^{O(k_2)}_{\mathrm{vector}}Z^{O(k_2)}_{\mathrm{hyper},\mathbf{sym}}Z^{O(k_2)}_{\mathrm{Fermi}},
\end{equation}
where
\begin{align}
Z^{O(k_2)}_{\mathrm{vector}}
&=\prod_{i=1}^{k_2/2}2\pi\eta\, \theta_1(2\epsilon_+)d\varphi_i
\prod_{1\leq i<j\leq k_2/2}\frac{\theta_1(\pm\varphi_i\pm\varphi_j)\theta(2\epsilon_+\pm\varphi_i\pm\varphi_j)}{\eta^8},\\
Z^{O(k_2)}_{\mathrm{hyper},\mathbf{sym}}
&=
\frac{\eta^{k_2}}{\theta^{\frac{k_2}{2}}_1(\epsilon_1)\theta_1^{\frac{k_2}{2}}(\epsilon_2)}\prod_{i=1}^{k_2/2}\frac{\eta^4}{\theta_1(\epsilon_1\pm2\varphi_i)\theta_1(\epsilon_2\pm2\varphi_i)} \prod_{i<j}\frac{\eta^8}{\theta_1(\epsilon_1\pm\varphi_i\pm\varphi_j)\theta(\epsilon_2\pm\varphi_i\pm\varphi_j)}, \nonumber\\
Z^{O(k_2)}_{\mathrm{Fermi}}
&=\prod_{i=1}^{k_2/2}\prod_{j=1}^{4}\frac{\theta_1(\pm\varphi_i\pm m_j)}{\eta^4}.\nonumber
\end{align}
Again we used the fact that the $k_2$-dimensional $O(k_2)$ charge vectors are given by
\begin{align}
\rho &=(0,\cdots,0,1,0,\cdots,0) && \text{for the defining representation},\\
\rho&=(0,\cdots,0,2,0,\cdots,0)\ \ \mathrm{or} \nn\\
&\ \ \ \ (0,\cdots,0,1,0,\cdots,0,1,0,\cdots,0) \nn
&& \text{for the adjoint representation},\\
\rho&=(0,\cdots,0,1,0,\cdots,0,1,0,\cdots,0)  && \text{for the antisymmetric representation}. \nn
\end{align}
$O(k_2)$ allows discrete holonomies. All disconnected holonomy sectors are classified into $k_2$ by $k_2$ matrices having the following eigenvalues:

\begin{small}
\begin{align}
O(2p+1):\,&\mathrm{diag}(\pm\varphi_1,\cdots,\pm\varphi_p,0),\,\mathrm{diag}(\pm\varphi_1,\cdots,\pm\varphi_{p-1},\half,\frac{1+\tau}{2},\frac{\tau}{2}),\,\mathrm{diag}(\pm\varphi_1,\cdots,\pm\varphi_p,\frac{\tau}{2}),\nn\\
&\mathrm{diag}(\pm\varphi_1,\cdots,\pm\varphi_{p-1},\half,\frac{1+\tau}{2},0),\,\mathrm{diag}(\pm\varphi_1,\cdots,\pm\varphi_p,\frac{1}{2}),\,\mathrm{diag}(\pm\varphi_1,\cdots,\pm\varphi_p,\frac{1+\tau}{2}),\nn\\
&\mathrm{diag}(\pm\varphi_1,\cdots,\pm\varphi_{p-1},0,\frac{\tau}{2},0),\,\mathrm{diag}(\pm\varphi_1,\cdots,\pm\varphi_{p-1},\frac{\tau}{2},\frac{1+\tau}{2},0)\nn\\
O(2p):\,&\mathrm{diag}(\pm\varphi_1,\cdots,\pm\varphi_p),\,\mathrm{diag}(\pm\varphi_1,\cdots,\pm\varphi_{p-2},0,\half,\frac{1+\tau}{2},\frac{\tau}{2}),\,\mathrm{diag}(\pm\varphi_1,\cdots,\pm\varphi_{p-1},0,\frac{\tau}{2}),\nn\\
&\mathrm{diag}(\pm\varphi_1,\cdots,\pm\varphi_{p-1},\half,\frac{1+\tau}{2}),\,\mathrm{diag}(\pm\varphi_1,\cdots,\pm\varphi_{p-1},0,\frac{1}{2}),\,\mathrm{diag}(\pm\varphi_1,\cdots,\pm\varphi_{p-1},\frac{\tau}{2},\half)\nn\\
&\mathrm{diag}(\pm\varphi_1,\cdots,\pm\varphi_{p-1},\frac{\tau}{2},\frac{1+\tau}{2}),\,\mathrm{diag}(\pm\varphi_1,\cdots,\pm\varphi_{p-1},0,\frac{1+\tau}{2}).
\end{align}
\end{small}

\noindent
We need to replace $(\rho \cdot \varphi)$'s with their proper value which involves discrete holonomies and sum over all distinct sectors. We will distinguish them by the superscript $d$, i.e., $Z^{O(k_2),d}$.

\subsubsection*{The bifundamental contribution}
The contribution from bifundamentals of $Sp(k_1)\times O(k_2)$ is $Z^{\mathrm{bif},d}=Z^{\mathrm{bif},d}_{\mathrm{twisted}}Z^{\mathrm{bif},d}_{\mathrm{Fermi}}$, with
\begin{equation}
\begin{split}
Z^{\mathrm{bif}}_{\mathrm{twisted}}&=
\prod_{i=1}^{k_1}\prod_{j=1}^{k_2/2}
\frac{\eta^4}{\theta_1(-\epsilon_+\pm\phi_i\pm\varphi_j)},\\
Z^{\mathrm{bif}}_{\mathrm{Fermi}}&=\prod_{i=1}^{k_1}\prod_{j=1}^{k_2/2}
\frac{\theta_1(\epsilon_-\pm\phi_i\pm\varphi_j)}{\eta^4}.
\end{split}
\end{equation}
The final $Sp(k_1) \times O(k_2)$ 1-loop determinant in a discrete holonomy sector $d$ is given by
\begin{align}
Z^{Sp(k_1)\times O(k_2), d}_{\rm 1-loop}  = Z^{Sp(k_1)}Z^{O(k_2),d} (Z^{\mathrm{bif},d})^2.
\end{align}
\begin{table}
\centering
\begin{tabular}{c|c|c}
Lie Algebra & Rep & Dynkin label \\
\hline
$Sp(k)$ & {\bf fund} & $(1,0,\cdots,0)$ \\
& {\bf adj} & $(2,0,\cdots,0)$\\
& {\bf anti} & $(0,1,0,\cdots,0)$\\
\hline
$O(k)$ & {\bf fund} & $(1,0,\cdots,0)$\\
& {\bf adj} & $(0,1,0,\cdots,0)$\\
& {\bf sym} & $(2,0,\cdots,0)$
\end{tabular}
\caption{\label{table:reps}The Dynkin labels of representations.}
\end{table}

\subsection{Matrix integral}
\subsubsection*{Pure $Sp(k_1)$ part}
First consider $Z^{Sp(k_1)}=Z^{Sp(k_1)}_{\mathrm{vector}}Z^{Sp(k_1)}_{\mathrm{hyper},\mathbf{anti}}Z^{Sp(k_1)}_{\mathrm{hyper}}$,
\begin{equation}
\begin{split}
Z^{Sp(k_1)}=&\prod_{i=1}^{k_1}2\pi\eta\theta_1(2\epsilon_+)d\phi_i
\frac{\theta_1(\pm2\phi_i)\theta_1(2\epsilon_+\pm2\phi_i)}{\eta^4}
\prod_{i=1}^{k_1}\prod_{j=1}^{4}\frac{\eta^4}{\theta_1(\epsilon_+\pm\phi_i\pm \mu_j)}\\
&\times\prod_{1\leq i<j\leq k_1}\frac{\theta_1(\pm\phi_i\pm\phi_j)\theta_1(2\epsilon_+\pm\phi_i\pm\phi_j)}{\theta_1(\epsilon_++\epsilon_-\pm\phi_i\pm\phi_j)\theta_1(\epsilon_+-\epsilon_-\pm\phi_i\pm\phi_j)}.
\end{split}
\end{equation}
Define the eigenvalue density and the profile function as
\begin{equation}
\label{eq:profile-so8}
\begin{split}
\rho(z)=&\e_1\e_2\sum_i\left(\delta(z-\phi_i)+\delta(z+\phi_i)\right),\\
f(z)=&-2\rho(z)+\sum_l(|z-\mu_l|+|z+\mu_l|).
\end{split}
\end{equation}
Under the thermodynamic limit $\e_1=-\e_2=\hbar$ and $\hbar\rightarrow 0$, we have the following expansion to leading order in $\hbar$,
\begin{equation}
Z^{Sp(k_1)} \sim \exp\left(-\frac{1}{\e_1\e_2}\CF^{Sp(k_1)}_0\right)=\exp\left(\frac{1}{\hbar^2}\CF^{Sp(k_1)}_0\right),
\end{equation}
and
\begin{equation} \label{eq:Fsp}
\begin{split}
\CF^{Sp(k_1)}_0
=&-\half\dashint dzdz'\rho(z)\rho(z')\partial_z^2\partial_{z'}^2\gamma_0(z-z')
-\half\int dz\rho(z)\partial^2_z\gamma_0(2z)\\
&+\int dz\rho(z)\sum_{l=1}^4\left(\partial^2_x\gamma_0(z+\mu_l)+\partial^2_z\gamma_0(z-\mu_l)\right).
\end{split}
\end{equation}
The function $\gamma_0(x)$ is the leading term in the expansion of the elliptic multiple Gamma function
\begin{equation}
	\gamma(z;\hbar) = \sum_{g=0}^{\infty} \hbar^{2g-2} \gamma_g(z),
\end{equation}
and in the following we will be needing the following property: $\gamma_0''(z) = \ln \theta_1(z)$. Applied to our present situation this gives
\begin{equation}
\begin{split}
\partial^2_z\gamma_0(z-z')=&\log\theta(z-z'),\\
\partial^2_z\partial_{z'}^2\gamma_0(z-z')=&-\frac{\theta'(z-z')^2-\theta(z-z')\theta''(z-z')}{\theta(z-z')^2}.
\end{split}
\end{equation}
We can now rewrite equation (\ref{eq:Fsp}) in terms of $f(x)$ which gives:
\begin{equation}
	\mathcal{F}_0^{Sp(k_1)} = -\frac{1}{8} \dashint dx dy f''(z)f''(z') \gamma_0(z-z') + \frac{1}{4}\int dz f''(z) \gamma_0(2z) + \mathcal{O}(\hbar^2).
\end{equation}

\subsubsection*{Pure $O(k_2)$ part}

First consider $k_2=2p$ even. 
\begin{equation}
\begin{split}
Z^{O(2p)}
&=\frac{\eta^{2p}}{\theta^{p}_1(\epsilon_1)\theta_1^{p}(\epsilon_2)}\prod_{i=1}^{p}2\pi\eta\theta_1(2\epsilon_+)d\varphi_i
\prod_{1\leq i<j\leq p}\frac{\theta_1(\pm\varphi_i\pm\varphi_j)\theta(2\epsilon_+\pm\varphi_i\pm\varphi_j)}{\theta_1(\epsilon_1\pm\varphi_i\pm\varphi_j)\theta(\epsilon_2\pm\varphi_i\pm\varphi_j)}\\
&\times
\prod_{i=1}^{p}\frac{\eta^4}{\theta_1(\epsilon_1\pm2\varphi_i)\theta_1(\epsilon_2\pm2\varphi_i)}\prod_{i=1}^{p}\prod_{j=1}^{4}\frac{\theta_1(\pm\varphi_i\pm \mu_j)}{\eta^4}.
\end{split}
\end{equation}
There are eight disconneted holonomy sectors and we need to replace $\varphi$'s with the correct holonomies. Notice that it is simpler if we group up holonomy sectors by the number of their continuous holonomies. For example, $\mathrm{diag}(\pm\varphi_1,\cdots,\pm\varphi_p)$ should be grouped with $\mathrm{diag}(\pm\varphi_1,\cdots,\pm\varphi_p,0,\half,\frac{1+\tau}{2},\frac{\tau}{2})$ although the first contribution is from holonomies of $O(2p)$ and the other one from holonomies of $O(2p+4)$. This is not a problem since we sum over all possible values of $p$.

We will be explicit here, $\log Z^{O(2p)}$ contains three terms. Under the limit $\e_1=-\e_2=\hbar$ and $\hbar\rightarrow 0$, the first one is
\begin{equation}
\begin{split}
&\log\prod_{1\leq i<j\leq p}\frac{\theta_1(\pm\varphi_i\pm\varphi_j)\theta(2\epsilon_+\pm\varphi_i\pm\varphi_j)}{\theta_1(\epsilon_1\pm\varphi_i\pm\varphi_j)\theta(\epsilon_2\pm\varphi_i\pm\varphi_j)}\\
=&\sum_{1\leq i<j\leq p}\hbar^2
\frac{\theta_1'(\pm\varphi_i\pm\varphi_j)^2-\theta_1(\pm\varphi_i\pm\varphi_j)\theta''_1(\pm\varphi_i\pm\varphi_j)}{\theta_1(\pm\varphi_i\pm\varphi_j)^2}+\CO(\hbar^4)\\
=&-\frac{1}{\hbar^2}\half\dashint dzdz'\vr(z)\vr(z')
\partial^2_z\partial^2_{z'}\gamma_0(z-z')+\CO(1).
\end{split}
\end{equation}
The second term is,
\begin{align}
\log\prod_{i=1}^{p}\frac{1}{\theta_1(\epsilon_1\pm2\varphi_i)\theta_1(\epsilon_2\pm2\varphi_i)}
&=-\sum_{i=1}^p\log\theta_1(\pm2\varphi_i)^2+\CO(\hbar^2)\nonumber\\
&=\frac{1}{\hbar^2}\half\dashint dz\vr(z)\partial^2_z\gamma_0(2z)+\CO(1).
\end{align}
The third term is,
\begin{equation}
\log\prod_{i=1}^{p}\prod_{j=1}^{4}\theta_1(\pm\varphi_i\pm \mu_j)
=-\frac{1}{\hbar^2}\int dz\vr(z)\sum_{l=1}^4\left(\partial^2_z\gamma_0(z+\mu_l)+\partial^2_z\gamma_0(z-\mu_l)\right).
\end{equation}
The contribution $\CF^{O(2p),1}_0$  of the continuous holonomy sector is given by
\begin{equation}
\begin{split}
\CF^{O(2p,1)}_0
=&-\half\dashint dzdz'\vr(z)\vr(z')\partial_z^2\partial_{z'}^2\gamma_0(z-z')
+\half\int dz\vr(z)\partial^2_z\gamma_0(2z)\\
&-\int dz\vr(z)\sum_{l=1}^4\left(\partial^2_z\gamma_0(z+\mu_l)+\partial^2_z\gamma_0(z-\mu_l)\right),
\end{split}
\end{equation}
with
\begin{equation}
\label{eq:profile-Estr}
\vr(z)=\e_1\e_2\sum_i(\delta(z-\varphi_i)+\delta(z+\varphi_i)).
\end{equation}
Now consider the effect of other holomony sectors. For example for $\varphi=\mathrm{diag}(\pm\varphi_1,\cdots,\pm\varphi_p,0,\frac{\tau}{2})$, the contribution from $Z^{O(2p)}_{\mathrm{Fermi}}$ is
\begin{equation}
Z^{O(2p)}_{\mathrm{Fermi}}=\left(\prod^p_{i=1}\prod_{j=1}^4\frac{\theta_1(\pm\varphi_i\pm \mu_j)}{\eta^4}\right)\left(\prod_{j=1}^4\frac{\theta_1(\pm \mu_j)}{\eta^2}\frac{\theta_1(\frac{\tau}{2}\pm \mu_j)}{\eta^2}\right).
\end{equation}
The discrete holonomies add only $\vr$ independent terms to $\CF_0^{O(2p)}$ hence can be omitted. Similar for the
\begin{equation}
\prod_{i=1}^{p+1}\frac{\eta^4}{\theta_1(\epsilon_1\pm2\varphi_i)\theta_1(\epsilon_2\pm2\varphi_i)}
\end{equation}
term where discrete holonomies have no effect. However, we have to be careful on the crossing terms,
\begin{small}
\begin{equation}
\begin{split}
&\prod_{1\leq i<j\leq p+1}\frac{\theta_1(\pm\varphi_i\pm\varphi_j)\theta_1(2\epsilon_+\pm\varphi_i\pm\varphi_j)}{\theta_1(\epsilon_1\pm\varphi_i\pm\varphi_j)\theta_1(\epsilon_2\pm\varphi_i\pm\varphi_j)}\\
=&\left(\prod_{1\leq i<j\leq p}\frac{\theta_1(\pm\varphi_i\pm\varphi_j)\theta_1(2\epsilon_+\pm\varphi_i\pm\varphi_j)}{\theta_1(\epsilon_1\pm\varphi_i\pm\varphi_j)\theta_1(\epsilon_2\pm\varphi_i\pm\varphi_j)}\right)\prod_{j=2p+1}^{2p+2}\left(\prod_{1\leq i\leq p}\frac{\theta_1(\pm\varphi_i+\varphi_j)\theta_1(2\epsilon_+\pm\varphi_i+\varphi_j)}{\theta_1(\epsilon_1\pm\varphi_i+\varphi_j)\theta_1(\epsilon_2\pm\varphi_i+\varphi_j)}\right)\times
\cdots,
\end{split}
\end{equation}
\end{small}
where we omitted terms which don't depend on $\varphi_i$ with $1\leq i\leq p$ and $\varphi_{2p+1}$ and $\varphi_{2p+2}$ are the two discrete holonomies. In this case $\varphi_{2p+1}=0$ and $\varphi_{2p+2}=\frac{\tau}{2}$. One can easily generalize this argument to the other five holonomy sectors with rank $2p+2$.
Under the thermodynamic limit, the crossing term is
\begin{equation}
\begin{split}
&\log\sum_{j=2p+1}^{2p+2}\sum_{1\leq i\leq p}\frac{\theta_1(\pm\varphi_i+\varphi_j)\theta_1(2\epsilon_+\pm\varphi_i+\varphi_j)}{\theta_1(\epsilon_1\pm\varphi_i+\varphi_j)\theta_1(\epsilon_2\pm\varphi_i+\varphi_j)}\\
=&\sum_{j=2p+1}^{2p+2}\sum_{1\leq i\leq p}\hbar^2
\frac{\theta_1'(\pm\varphi_i+\varphi_j)^2-\theta_1(\pm\varphi_i+\varphi_j)\theta''_1(\pm\varphi_i+\varphi_j)}{\theta_1(\pm\varphi_i+\varphi_j)^2}+\CO(\hbar^4)\\
=&-\int dz \vr(z)\left[\partial^2_z\partial^2_{z'}\left.\gamma_0(z-z')\right|_{z'=-\varphi_{2p+1}}+\partial^2_z\partial^2_{z'}\left.\gamma_0(z-z')\right|_{z'=-\varphi_{2p+2}}\right]+\CO(\hbar^2),
\end{split}
\end{equation}
which is apparently higher order. Therefore the leading contribution to $\CF$ from this discrete holonomy sector of rank $2p+2$ is the same as the one of continuous holonomy of rank $2p$. For the holonomy sector with rank $2p+4$ we have
\begin{small}
\begin{equation}
\begin{split}
&\prod_{1\leq i<j\leq p+1}\frac{\theta_1(\pm\varphi_i\pm\varphi_j)\theta(2\epsilon_+\pm\varphi_i\pm\varphi_j)}{\theta_1(\epsilon_1\pm\varphi_i\pm\varphi_j)\theta(\epsilon_2\pm\varphi_i\pm\varphi_j)}\\
=&\left(\prod_{1\leq i<j\leq p}\frac{\theta_1(\pm\varphi_i\pm\varphi_j)\theta(2\epsilon_+\pm\varphi_i\pm\varphi_j)}{\theta_1(\epsilon_1\pm\varphi_i\pm\varphi_j)\theta(\epsilon_2\pm\varphi_i\pm\varphi_j)}\right)\prod_{j=2p+1}^{2p+4}\left(\prod_{1\leq i\leq p}\frac{\theta_1(\pm\varphi_i+\varphi_j)\theta(2\epsilon_+\pm\varphi_i+\varphi_j)}{\theta_1(\epsilon_1\pm\varphi_i+\varphi_j)\theta(\epsilon_2\pm\varphi_i+\varphi_j)}\right)\times
\cdots,
\end{split}
\end{equation}
\end{small}
And again the crossing terms do not contribute to the leading order of $\CF$. Therefore six holonomy sectors with rank $2p+2$ and one holonomy sector with rank $2p+4$ have the same leading order as the continuous sector with rank $2p$. Altogether, we can summarize the contributions of the $O(k_2)$-node as follows
\begin{eqnarray}
	\mathcal{F}^{O(k_2)}_0 & = & - \frac{1}{2} \dashint \varrho(z) \varrho(z') \partial_z^2 \partial_{z'}^2 \gamma_0(z-z') + \half \int dz \varrho(z) \partial_z^2 \gamma_0(2z) \nonumber \\
	~ & ~ & - \int dz \varrho(z) \sum_{l=1}^4 (\partial_z^2 \gamma_0(z+\mu_l) + \partial_z^2\gamma_0(z-\mu_l)).
\end{eqnarray}
Defining $g(z) = -2 \varrho(z)$ then gives
\begin{eqnarray}
	\mathcal{F}^{O(k_2)}_0 & = &- \frac{1}{8} \dashint dz dz' g''(z) g''(z') \gamma_0(z-z') - \frac{1}{4} \int dx g''(z) \gamma_0(2z) \nonumber \\
	~ & ~ & -\int dx \varrho''(z) \sum_{l=1}^4 (\gamma_0(z+\mu_l) + \gamma_0(z-\mu_l)).
\end{eqnarray}

%

\subsubsection*{Bifundamental part}
The bifundamental contribution is,
\begin{equation}
\begin{split}
(Z^{\mathrm{bif}})^2&=
\prod_{i=1}^{k_1}\prod_{j=1}^{p}
\frac{\theta^2_1(\epsilon_-\pm\phi_i\pm\varphi_j)}{\theta^2_1(-\epsilon_+\pm\phi_i\pm\varphi_j)}.
\end{split}
\end{equation}
where $\phi_i$ and $\varphi_j$ denote eigenvalues of $Sp(k_1)$ and $O(k_2)$ holonomies, respectively. For discrete $O(k_2)$ holonomy sectors, some $\varphi_j$ can represent either $0$ or the half-period points on $T^2$, i.e., $\frac{1}{2}$, $\frac{1+\tau}{2}$, $\frac{\tau}{2}$. For the continuous $O(k_2)$ holonomy sector, 
\begin{align}
&\log\prod_{i=1}^{k_1}\prod_{j=1}^{p}
\frac{\theta^2_1(\epsilon_-\pm\phi_i\pm\varphi_j)}{\theta^2_1(-\epsilon_+\pm\phi_i\pm\varphi_j)}\\
=&2\hbar\sum_{i=1}^{k_1}\sum_{j=1}^{p}\frac{\theta_1'( \pm \phi_i \pm \varphi_j)}{\theta_1(\pm\phi_i \pm \varphi_j)}
 -\hbar^2\sum_{i=1}^{k_1}\sum_{j=1}^{p} \frac{\theta_1'(\pm\varphi_i+\varphi_j)^2-\theta_1(\pm\varphi_i+\varphi_j)\theta''_1(\pm\varphi_i+\varphi_j)}{\theta_1(\pm\varphi_i+\varphi_j)^2}+\CO(\hbar^3).\nn
\end{align}
The leading term vanishes by taking the sum over the $\pm$ sectors, since $\theta_1$ is odd and  its derivative is even. Using the functions $f(z)$ and $g(z')$,
the bifundamental contribution is given by
\begin{equation}
	\mathcal{F}_0^{\textrm{bif}} = \frac{1}{4} \dashint dz dz' f''(z) g''(z') \gamma_0(z-z') + \int dz \varrho''(z) \sum_{l=1}^4 (\gamma_0(z+\mu_l) + \gamma_0(z-\mu_l)).
\end{equation}
We see that the second term on the right hand side is exactly equal to the negative of the last term of $\mathcal{F}_0^{O(k_2)}$. Thus we conclude that this term cancels in the overall product of gauge and matter contributions.

\subsection{The full partition function}
We now gather all results from the previous subsection and combine everything into an expression for the full partition function
\begin{equation}
Z=Z^{pert}\left(1+\sum_{k_1, k_2\geq0}q_1^{k_1}q_0^{k_2}\sum_d Z^{Sp(k_1)\times O(k_2),d}\right).
\end{equation}
In the above $Z^{pert}$ is the perturbative contribution to the partition function and we will henceforth leave it unspecified as it will play no role in further discussions. Using the fact that
\begin{equation}
\int dx x^2 f''(z) = 4 \sum_l \mu_l^2 + 8 \hbar^2 k_1,
\end{equation}
and
\begin{equation}
\int dx x^2 g''(z)= 8 \hbar^2 k_2,
\end{equation}
we have
\begin{equation}
\begin{split}
Z=&Z^{pert}\int[d f''(z)][d g''(z)]e^{\frac{\CF^{\textrm{full}}_0}{\hbar^2}}\left(
1+q_0^2\sum_d e^{\CF^{d,(1)}}+q_0^4e^{\CF^{8,(1)}}\right).
\end{split}
\end{equation}
Note that, as stated before, contributions from discrete holonomy sectors, captured by $\mathcal{F}^{d,(1)}$ and $\mathcal{F}^{8,(1)}$, are of higher order in $\hbar$ and can be treated as perturbations to the leading order contribution $\mathcal{F}^{\textrm{full}}_0$ of continuous holonomies. The leading order contribution is given by
\begin{eqnarray}
	\mathcal{F}^{\textrm{full}}_1 & = & \frac{2\pi i \tau_{b,1}}{8} \int dz z^2 f''(z) - \frac{1}{8} \dashint dz dz' f''(z) f''(z') \gamma_0(z-z') + \frac{1}{4}\int dz f''(z) \gamma_0(2z) \nonumber \\
	~ & ~ & \frac{2\pi i \tau_{b,0}}{8} \int dz z^2 g''(z) - \frac{1}{8} \dashint dz dz' g''(z) g''(z') \gamma_0(z-z') - \frac{1}{4} \int dz g''(z) \gamma_0(2z) \nonumber \\
	~ & ~ & + \frac{1}{4} \dashint dz dz' f''(z) g''(z') \gamma_0(z-z'), 
\end{eqnarray}
where we have defined $\log(q_0) = 2\pi i \tau_{b,0}$ and $\log(q_1) = 2\pi i \tau_{b,1}$. Variation with respect to $f''(z)$ and $g''(z)$ then gives the following saddle point equations:\footnote{From now on we will denote $\mathcal{F}_0^{\textrm{full}}$ simply as $\mathcal{F}_0$.}
\begin{eqnarray}
	\frac{\delta \mathcal{F}_0}{\delta f''(z)} & = & \frac{2\pi i \tau_{b,1}}{8}z^2 - \frac{1}{4}\int dz' f''(z') \gamma_0(z-z') + \frac{1}{4}\gamma_0(2z) \nonumber \\
	~ & ~&  + \frac{1}{4} \int dz' g''(z') \gamma_0(z-z') = 0, \nonumber \\
	\frac{\delta \mathcal{F}_0}{\delta g''(z)} & = & \frac{2\pi i \tau_{b,0}}{8}z^2 - \frac{1}{4}\int dz'  g''(z') \gamma_0(z-z') - \frac{1}{4}\gamma_0(2z) \nonumber \\
	~ & ~ &  + \frac{1}{4} \int dy f''(z') \gamma_0(z-z') \nonumber \\
	~ & = & 0.
\end{eqnarray}
Multiplying by $4$ and taking $2$ derivatives with respect to $z$ gives:\footnote{$y^{\pm}_i$ are defined by $y_i^{\pm}(z) \equiv y_i(z\pm i\epsilon)$.}
\begin{eqnarray}
	y_0^+(z) y_0^-(z) & = & \mathcal{P}_0 y_1(z)^2 \label{eq:y0}\\
	y_1^+(z) y_1^-(z) & = & \mathcal{P}_1 y_0(z)^2 \label{eq:y1},
\end{eqnarray}
where 
\begin{eqnarray}
	\mathcal{P}_0 & = & \theta_1(2z)^{-4} q_0, \nonumber \\
	y_0 & = & \exp \half \int dz' g''(z') \log \theta_1(z-z'), \nonumber \\
	\mathcal{P}_1 & = & \theta_1(2z)^4 q_1 , \nonumber \\
	y_1 & = & \exp \half \int dz' f''(z') \log \theta_1(z-z').
\end{eqnarray}
Equation (\ref{eq:y0}) and (\ref{eq:y1}) can then be rewritten as transformations
\begin{eqnarray} \label{eq:ri}
	r_0 & : & y_0 \mapsto \mathcal{P}_0 \frac{y_1^2}{y_0}, \nonumber \\
	r_1 & : & y_1 \mapsto \mathcal{P}_1 \frac{y_0^2}{y_1}
\end{eqnarray}
upon crossing cuts on the $z$-plane which due to periodicity properties of the $\theta_1$ functions is actually compactified to a torus $\mathbb{T}_{\tau_f} = \mathbb{C}/(\mathbb{Z} \oplus \tau_f \mathbb{Z})$. Equations (\ref{eq:ri}) then describe a two-sheeted covering of this torus with $y_1$ being a coordinate on one sheet with cuts at $I_{0,\mu_l}$ and $I_{0,-\mu_l}$ for $l=1,2,3,4$. On the other hand, $y_0$ is a coordinate on the other sheet with only one cute, namely $I_{1,0}$. Then $r_i$ for $i=0,1$ can be interpreted as Weyl reflections of the affine $\widehat{A}_1$ quiver on these sheets. To make this clear, we rewrite equations (\ref{eq:ri}) as follows
\begin{equation} \label{eq:iWeyl}
	r_i : y_i \mapsto \mathcal{P}_i y_i \prod_j y_j^{-C_{ij}},
\end{equation}
where the product over $j$ is the one over all nodes of the quiver and $C_{ij}$ is the Cartan matrix
\begin{equation}
	C_{ij} = \left(\begin{array}{cc}2 & -2 \\ -2 & 2\end{array}\right).
\end{equation}
Equation \eqref{eq:iWeyl} is referred to in \cite{Nekrasov:2012xe} as an iWeyl reflection. At this point, we want to highlight a symmetry enjoyed by equations (\ref{eq:iWeyl}). Notice that under the $\mathbb{Z}_2$ reflection $z \mapsto -z$, $y_1$ transforms as follows
\begin{eqnarray}
	y_1 & \mapsto & \exp \half \int dz' f''(z') \log \theta_1(-z - z') \nonumber \\
	~     & =           & \exp\left( - \half \int dz'' f''(-z'') \log \theta_1(-z + z'') \right) \nonumber \\
	~     & =           & \exp \left( - \half \int dz'' f''(z'') (\log \theta_1(z-z'') + i \pi)\right) \nonumber \\
	~     & =           & y_1^{-1},
\end{eqnarray}
where in the last two steps we have used that $f''(z'')$ is an even function and that $\int dz'' f''(z'')$ is an integer multiple of $4$. Similarly, we can show that 
\begin{equation}
	z \mapsto -z \quad \longrightarrow \quad y_0 \mapsto y_0^{-1}.
\end{equation}
Furthermore, adding the transformation $\mathcal{P}_i \mapsto {\mathcal{P}_i}^{-1}$, we see that the combined reflection
\begin{equation} \label{eq:Z2}
	z \mapsto -z, \quad \mathcal{P}_i \mapsto {\mathcal{P}_i}^{-1},
\end{equation}
is a symmetry of the saddle point equations (\ref{eq:iWeyl}). This observation is very crucial for the derivation of the spectral curve which we will be dealing with in detail in the next section. Before proceeding to the derivation of the spectral curve, we will comment on the more general story of $\mathcal{T}^A_{D_{4+n},N}$, namely $N$ M5 branes probing a $D_{4+n}$ singularity.

\subsection{Thermodynamic limit for general $\mathcal{T}^A_{D_{n},N}$ theory}

Let us first focus on the immediate generalization of one M5 brane probing a $D_{n}$ singularity with $n > 4$ and then proceed to the general picture.
\subsubsection*{One brane on $D_n$ ($N=1$)}
\begin{figure}
\centering
\includegraphics[scale=0.7]{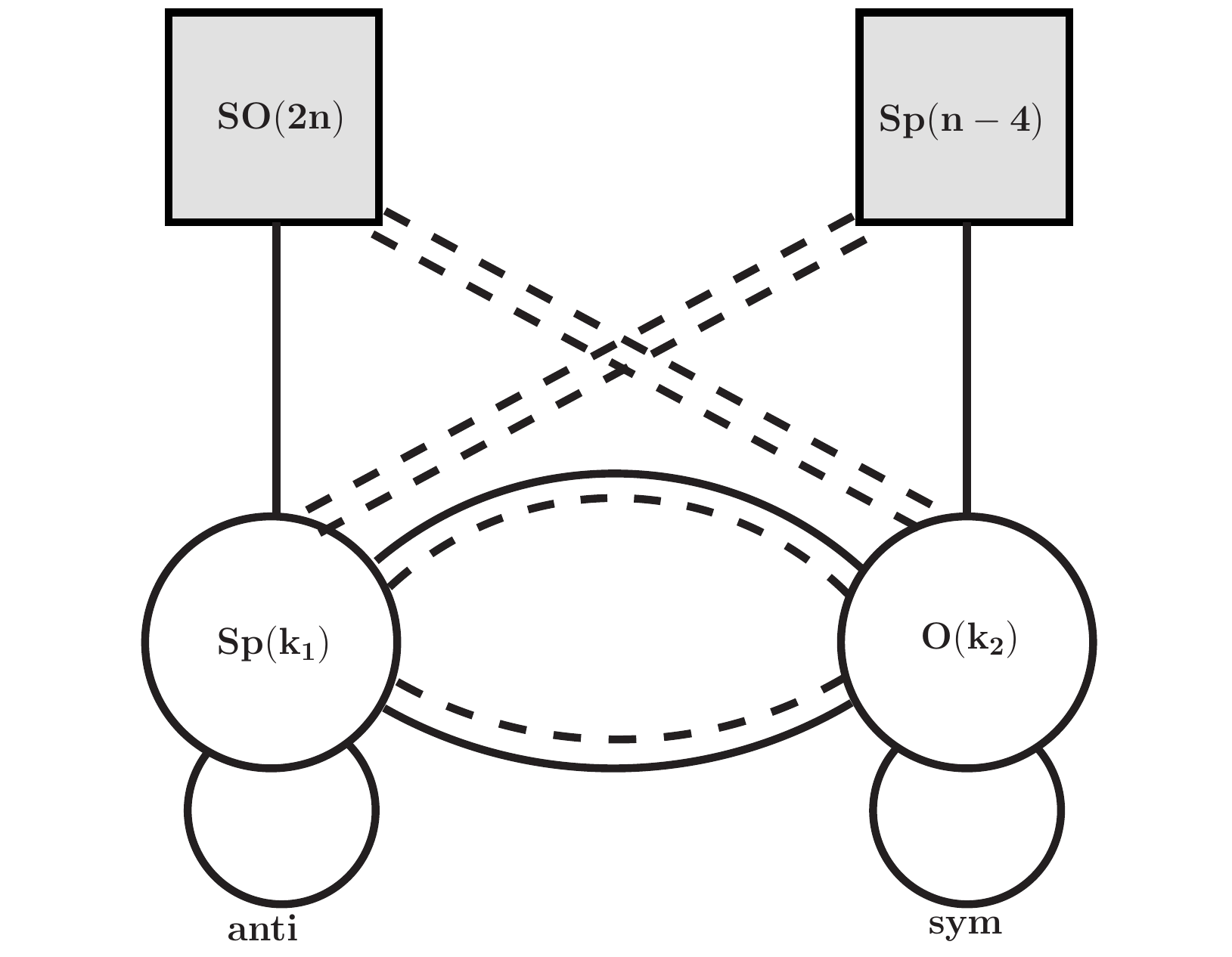}
\caption{\label{fig:quiverDnN1}Quiver for one instanton on $D_n$. }
\end{figure}

This adds one $Sp(n-4)$ flavor node in the quiver together with a bifundamental hyper of $Sp(n-4)\times O(k_2)$ and $2$ bifundamental Fermi of $Sp(n-4)\times Sp(k_1)$. The quiver is depicted in figure \ref{fig:quiverDnN1}.
Therefore, the total matter contribution to the pure $Sp(k_1)$ part of the partition function is given by
\begin{equation}
\prod_{i=1}^{k_1}\prod_{j=1}^n\frac{\eta^4}{\theta_1(\e_+\pm\phi_i\pm a_i)}
\prod_{i=1}^{k_1}\prod_{j=1}^{n-4}\frac{\theta_1(\pm\phi_i\pm b_j)}{\eta^4}.
\end{equation}
The total matter contribution to the pure $O(k_2)$ part of the partition function is
\begin{equation}
\prod_{i=1}^{k_2}\prod_{j=1}^{n}\frac{\theta_1(\pm\vr_i\pm a_j)}{\eta^4}\prod_{i=1}^{k_2}\prod_{j=1}^{n-4}\frac{\eta^4}{\theta_1(\e_+\pm\vr_i\pm b_j)}.
\end{equation}
We use $a$'s and $b$'s to denote fugacities of $SO(2n)$ and $Sp(n-4)$ flavor symmetry respectively. Therefore the free energy at thermodynamic limit of pure $Sp(k_1)$ part is
\begin{equation} \label{eq:FspDn}
\begin{split}
\CF^{Sp(k_1)}_0
\sim&-\half\dashint dzdz'\rho(z)\rho(z')\partial_z^2\partial_{z'}^2\gamma_0(z-z')
-\half\int dz\rho(z)\partial^2_z\gamma_0(2z)\\
&+\int dz\rho(z)\sum_{f=1}^n\left(\partial^2_z\gamma_0(z+a_f)+\partial^2_z\gamma_0(z-a_f)\right)\\
&-\int dz\rho(z)\sum_{f=1}^{n-4}\left(\partial^2_z\gamma_0(z+b_f)+\partial^2_z\gamma_0(z-b_f)\right).
\end{split}
\end{equation}
and the pure $O(k_2)$ part is
\begin{eqnarray}
	\mathcal{F}^{O(k_2)}_0 & = & - \frac{1}{2} \dashint \varrho(z) \varrho(z') \partial_z^2 \partial_{z'}^2 \gamma_0(z-z') + \half \int dz \varrho(z) \partial_z^2 \gamma_0(2z) \nonumber \\
	~ & ~ & - \int dz \varrho(z) \sum_{f=1}^n (\partial_z^2 \gamma_0(z+a_f) + \partial_z^2\gamma_0(z-a_f))\\
	~ & ~ & + \int dz \varrho(z) \sum_{f=1}^{n-4} (\partial_z^2 \gamma_0(z+b_f) + \partial_z^2\gamma_0(z-b_f)).
\end{eqnarray}
The bifundamental part remains the same. Defining
\begin{equation} \label{eq:fgn}
\begin{split}
f(z)=&-2\r(z)+\sum_{f=1}^n(|z-a_f|+|z+a_f|),\\
g(z)=&-2\vr(z)+\sum_{f=1}^{n-4}(|z-b_f|+|z+b_f|),
\end{split}
\end{equation}
the total free energy in thermodynamic limit is,
\begin{eqnarray}
	\mathcal{F}_0 & = & \frac{2\pi i \tau_{b,1}}{8} \int dz z^2 f''(z) - \frac{1}{8} \dashint dz dz' f''(z) f''(z') \gamma_0(z-z') + \frac{1}{4}\int dz f''(z) \gamma_0(2z) \nonumber \\
	~ & ~ & \frac{2\pi i \tau_{b,0}}{8} \int dz z^2 g''(z) - \frac{1}{8} \dashint dz dz' g''(z) g''(z') \gamma_0(z-z') - \frac{1}{4} \int dz g''(z) \gamma_0(2z) \nonumber \\
	~ & ~ & + \frac{1}{4} \dashint dz dz' f''(z) g''(z') \gamma_0(z-z').
\end{eqnarray}
The saddle-point equations remain the same as in (\ref{eq:iWeyl}) with the only difference that the $y_i$ functions are defined with the new $f(z)$ and $g(z)$ functions given in (\ref{eq:fgn}). In particular, the saddle-point equations will still enjoy the $\mathbb{Z}_2$ symmetry (\ref{eq:Z2}).

\subsubsection*{$N$ branes on $D_n$}
\begin{figure}
\centering
\includegraphics[scale=0.5]{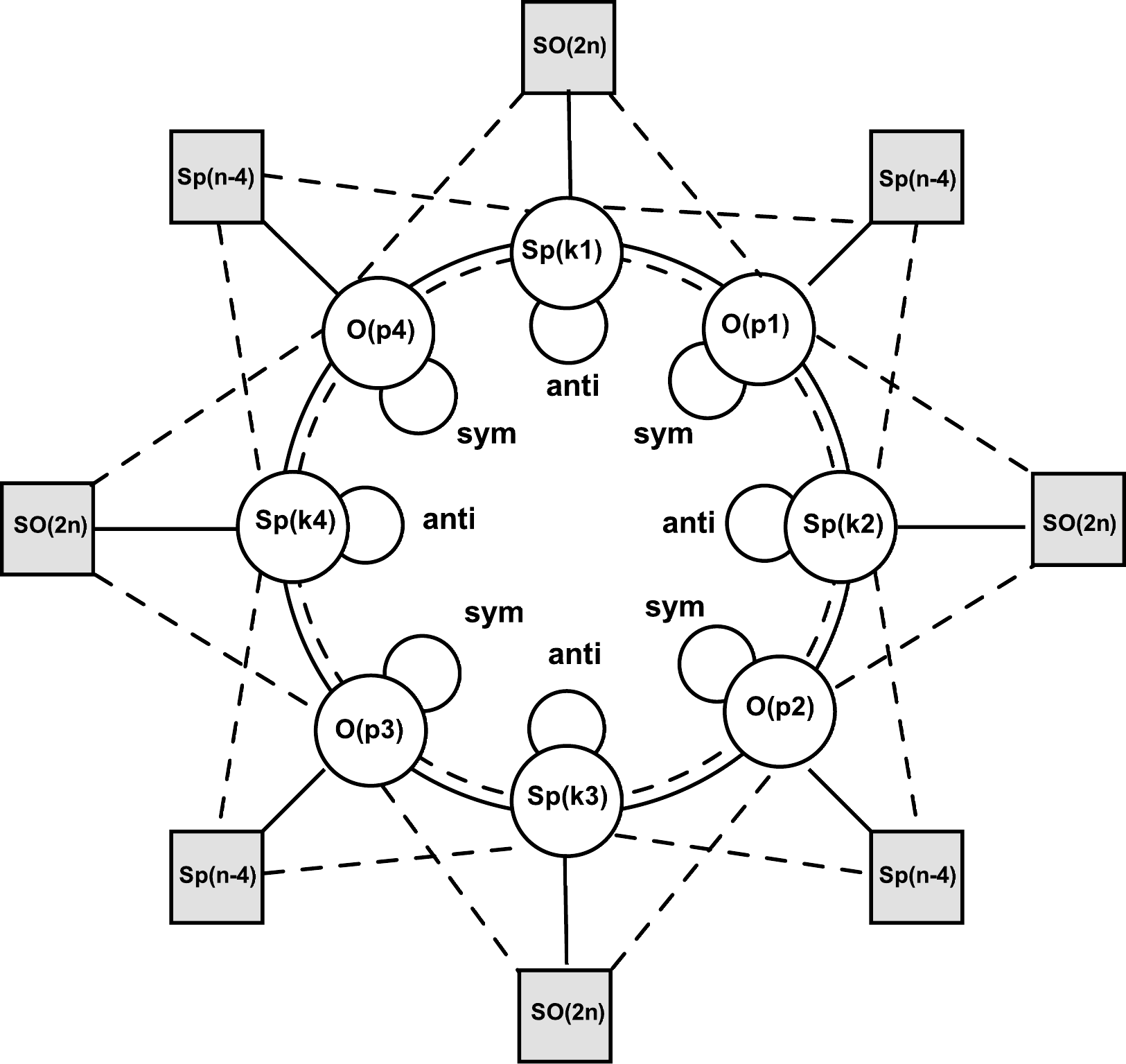}
\caption{\label{fig:quiverDnN}Quiver for $N$ branes on $D_n$.}
\end{figure}

We will have $2N$ gauge nodes alternating between $Sp$ groups and $O$ groups. The quiver diagram for $N=4$ is depicted in figure \ref{fig:quiverDnN}. We use $k_1, \cdots,k_N$ and $p_1,\cdots,p_N$ to denote the winding modes.
The total free energy is
\begin{equation}
\begin{split}
\CF_0=&\sum_{i=1}^N\left\{\frac{2\pi i \tau_{b,1}}{8} \int dz z^2 f_i''(z) - \frac{1}{8} \dashint dz dz' f_i''(z) f_i''(z') \gamma_0(z-z') + \frac{1}{4}\int dz f_i''(z) \gamma_0(2z)\right. \\
&\frac{2\pi i \tau_{b,0}}{8} \int dz z^2 g''_i(z) - \frac{1}{8} \dashint dz dz' g''_i(z) g''_i(z') \gamma_0(z-z') - \frac{1}{4} \int dz g''_i(z) \gamma_0(2z)\\
&\left. + \frac{1}{8} \dashint dz dz' f''_i(z) g''_i(z') \gamma_0(z-z')
+ \frac{1}{8} \dashint dz dz' g''_i(z) f_{i+1}''(z') \gamma_0(z-z')\right\},
\end{split}
\end{equation}
where $N+1$ is identified with $1$. $f_i$'s and $g_i$'s are
\begin{equation}
\label{eq:profile-general}
\begin{split}
f_i(z)=&-2\r_i(z)+\sum_{f=1}^n(|z-a_{i,f}|+|z+a_{i,f}|),\\
g_i(z)=&-2\vr_i(z)+\sum_{f=1}^{n-4}(|z-b_{i,f}|+|z+b_{i,f}|).
\end{split}
\end{equation}
The saddle-point equations are the same as (\ref{eq:iWeyl}) with $C_{ij}$ now replaced with a general affine $\widehat{A}_{2N-1}$ Cartan matrix. Notice that the $\mathbb{Z}_2$ reflection (\ref{eq:Z2}) is a symmetry of the equations.

\section{Spectral curves from saddle-point equations}
\label{sec:curve}

Let us construct the Seiberg-Witten curves of little string theories $\mathcal{T}^A_{D_4,N}$ and $\mathcal{T}^B_{D_4,N}$ engineered from M5/D5 branes probing an affine $D_4$ singularity. In the case of $\mathcal{T}^B_{D_4,N}$, the curve corresponds to the zero locus of the determinant of an $SO(8)$-bundle over $\widehat{\mathbb{T}}_f$ as was derived in \cite{Haghighat:2017vch}. Here, we want to see what the corresponding section for $\mathcal{T}^A_{D_4,N}$ looks like and subsequently compare the two spectral curves obtained. 

To begin with, we note that we need to construct a section which is invariant under the reflections (\ref{eq:iWeyl}). To this end, we define variables
\begin{equation}
	t_i(z) = \check{t}_i \frac{y_i(z)}{y_{i-1}(z)}, i=1,\ldots,2N,
\end{equation}
where we impose the periodicity condition $y_{2N} = y_0$ defining $y_i$ for all $i \in \mathbb{Z}$. Thus 
\begin{equation}
	\mathbf{t}(z) = (t_1(z),t_2(z),\ldots,t_{2N}(z))
\end{equation}
represents an element of the maximal torus of $SL(2N,\mathbb{C})$, i.e.
\begin{equation}
	\prod_{i=1}^{2N} t_i(z) = 1.
\end{equation}
The $\check{t}_i$ are defined by
\begin{equation}
	\check{t}_{i} = (\mathcal{P}_i \ldots \mathcal{P}_{2N-1})^{-1} (\mathcal{P}_1 \mathcal{P}_2^2 \ldots \mathcal{P}_{2N-1}^{2N-1})^{\frac{1}{2N}}, \quad i = 1, \ldots, 2N.
\end{equation}
Then it is easy to check that the following zero-set of the determinant of an $SL(2N,\mathbb{C})$ bundle over $\mathbb{T}_b$
\begin{equation} \label{eq:sA}
	s^A_{N}(w,\{w_l\};\tau_b) = \prod_{i=1}^{2N} \frac{\theta_1(w-w_l(z);\tau_b)}{\theta_1(w;\tau_b)} = 0, \quad t = e^{2\pi i w},~ t_i  = e^{2\pi i w_i(z)},
\end{equation}
is in fact invariant under the reflections (\ref{eq:iWeyl}) derived from the saddle-point equations. To derive this, one has to use periodicity properties of the $\theta_1$-function as reviewed in the appendix. Note that in a strict sense (\ref{eq:sA}) is the restriction of another section of a degree zero line bundle on $\mathbb{T}_b \times \widehat{\mathbb{T}}_f$ where $z$ is a coordinate on $\widehat{\mathbb{T}}_f$. The $\mathbb{Z}_2$ symmetry given in (\ref{eq:Z2}) acts then as follows
\begin{equation}
	t_i(z) \mapsto 1/t_i(z),
\end{equation}
which combined with (\ref{eq:sA}) immediately tells us that this symmetry lifts to $\mathbb{T}_b \times \widehat{\mathbb{T}}_f$:
\begin{equation}
	\mathbb{Z}_2 : \quad z \mapsto -z, \quad w \mapsto - w,
\end{equation}
which in turn tells us that the spectral curve (\ref{eq:sA}) lives in the K3 surface $(\mathbb{T}_b \times \widehat{\mathbb{T}}_f)/\mathbb{Z}_2$. More intuitively, we can think of this K3 surface as an elliptic fibration over $\mathbb{P}^1 = \widehat{\mathbb{T}}_f/\mathbb{Z}_2$ with fibers given by $\mathbb{T}_b$ everywhere except over the $4$ fixed-points under the $\mathbb{Z}_2$ in the base where they degenerate according to a type $I_0^*$ Kodaira singularity. The spectral curve can then be viewed as a $2N$-fold cover of the $\mathbb{P}^1$-base with restriction to the fiber given by (\ref{eq:sA}). 

Let us now turn to the spectral curve for $\mathcal{T}^B_{D_4,N}$. Adopting the notation $\theta_n(x\pm y) \equiv \theta_n(x+ y) \theta_n(x- y) $ for simplicity, the $SO(8)$-section found in \cite{Haghighat:2017vch} can be written as
\begin{equation}
\label{eq:sB}
s^{B} (w,\{w_l\};\tau_f) = \prod_{l=1}^4 \frac{\theta_1(w \pm w_l(z); \tau_f)}{\theta_1(w;\tau_f)^2}=0,
\end{equation}
where in this case $w$ is now a coordinate on $\widehat{\mathbb{T}}_f$ and $z$ a coordinate on $\mathbb{T}_b$. As there is no $\mathbb{Z}_2$ symemtry here, the spectral curve is a hypersurface inside $\widehat{\mathbb{T}}_f \times \mathbb{T}_b$. The Seiberg-Witten differential of both curves $s^A=0$ and $s^B=0$ takes the form
\begin{equation}
	\lambda_{\textrm{SW}} = z \frac{dX(w)}{Y(w)},
\end{equation}
where $X$ and $Y$ are Weierstrass coordinates whose precise definitions are given in Appendix \ref{sec:notation}.  Next, we want to expand the sections $s^A$ and $s^B$ in terms of Weierstrass coordinates $X(w;\tau)$ and $Y(w;\tau)$. The following identity \cite{Bertola:1999} can be utilized to manipulate the $SL(N+1)$ determinant line bundles, for which $\sum_{i=1}^{N+1} w_i = 0$ holds.
 \begin{align}
 \label{eq:decompose-identity}
\prod_{i=1}^{N+1} \frac{ \eta(\tau)^3 \, \theta_1(w- w_l;\tau)}{\theta_1(w_l;\tau) \theta_1(w;\tau)} &= \frac{(-1)^{N}}{2^{N-1} N!} \frac{\text{det}\left(\begin{matrix} 1 & \wp (w;\tau) & \wp' (w;\tau) & \cdots & \wp^{(N-1)}(w;\tau)\\
 1 & \wp (w_1;\tau) & \wp' (w_1;\tau) & \cdots & \wp^{(N-1)}(w_1;\tau)\\
 \vdots & \vdots & \vdots &  & \vdots \\
 1 & \wp (w_{N};\tau) & \wp' (w_{N};\tau) & \cdots & \wp^{(N-1)}(w_{N};\tau)\end{matrix}\right)}{\text{det}\left(\begin{matrix} 
 1 & \wp (w_1;\tau) & \wp' (w_1;\tau) & \cdots & \wp^{(N-2)}(w_1;\tau)\\
 \vdots & \vdots & \vdots &  & \vdots \\
 1 & \wp (w_{N};\tau) & \wp' (w_{N};\tau) & \cdots & \wp^{(N-2)}(w_{N};\tau)\end{matrix}\right)}.
 \end{align}
 
 \paragraph{SL(2)} Once we apply \eqref{eq:decompose-identity} to $s^A_1(w,\{w_i\};\tau_b)$ with $w_1 + w_2 = 0$, it is found that
\begin{align}
\label{eq:SU2-decompose}
	s^{A}_1 (w,\{w_1\};\tau_b) &=   \frac{\theta_1(w_1;\tau_b )^2\,  \wp (w_1;\tau_b) }{\eta(\tau_b)^6 } \cdot 1 -  \frac{\theta_1(w_1;\tau_b )^2}{4\eta(\tau_b)^6 } \cdot X (w;\tau_b) 
\end{align}
We want to express coefficients of the Weierstrass monomials using two $\hat{A}_1$ fundamental characters. Both of them are at level-1, being associated to the irreducible representations whose  highest weights are $\mathbf{g}^{[0]} =[0,0]$ and $\mathbf{g}^{[1]}=[+\frac{1}{2},-\frac{1}{2}]$:
\begin{align}
\Theta_{A_1}^{[0]}(w_1;\tau_b) = \theta_3(2w_1;2\tau_b),\quad \Theta_{A_1}^{[1]}(w_1;\tau_b) = \theta_2(2w_1;2\tau_b).
\end{align}
Note that both $\Theta_{A_a}^{[0]}$ as well as $\Theta_{A_1}^{[1]}$ are invariant under any combinations of Weyl reflections (\ref{eq:iWeyl}). Thus they must be invariant under the operation of crossing cuts in the $z$-plane and hence must be sections of degree zero line bundles on $\mathbb{T}_b$. We will make strong use of this observation in section \ref{sec:fibersections} to derive their concrete $z$-dependence. For now, let us proceed by noting the following identities
\begin{align}
\label{eq:identity}
\theta_1(w;\tau_b)^2   &= \theta_2(0;2\tau_b)\theta_3(2w;2\tau_b) -\theta_3(0;2\tau_b)\theta_2(2w;2\tau_b) \\
 \theta_4(w;\tau_b)^2 &=  \theta_3(0;2\tau_b)\theta_3(2w;2\tau_b)-\theta_2(0;2\tau_b)\theta_2(2w;2\tau_b),\nonumber
 \end{align}
which applied to the coefficients  in \eqref{eq:SU2-decompose} allows us to write them as follows: (with $\theta_n \equiv \theta_n(0;\tau_b)$ being understood)
\begin{align}
\frac{\theta_1(w_1;\tau_b)^2}{\eta(\tau_b)^6} &= \frac{\theta_2(0;2\tau_b)}{\eta(\tau_b)^6} \cdot \Theta_{A_1}^{[0]}(w_1;\tau_b)- \frac{\theta_3(0;2\tau_b)}{\eta(\tau_b)^6}  \cdot \Theta_{A_1}^{[1]}(w_1;\tau_b), \label{eq:theta1ch} \\
\frac{\theta_1(w_1;\tau_b)^2\,\wp(w_1;\tau_b)}{\eta(\tau_b)^6} &= \frac{3 \theta_2^2\theta_3^2 \theta_3(0;2\tau_b) -\theta_2(0;2\tau_b) \big(\theta_2^4 + \theta_3^4\big)}{12\, \eta (\tau_b)^6} \cdot \Theta_{A_1}^{[0]}(w_1;\tau_b)\nonumber\\
&-\frac{3 \theta_2^2\theta_3^2 \theta_2(0;2\tau_b) -\theta_3(0;2\tau_b) \big(\theta_2^4 + \theta_3^4\big)}{12\, \eta (\tau_b)^6} \cdot \Theta_{A_1}^{[1]}(w_1;\tau_b). \label{eq:theta4ch}
\end{align}

%
\paragraph{SL(4)} Similarly, the $SL(4)$ section $s^A_2(w,\{w_i\};\tau_b)$ with $\sum_{i=1}^4 w_4 = 0$ is decomposed into
\begin{align} \label{eq:sl4}
	s^{A_3}(w,\{w_l\};\tau_b)  &= a_0(\{w_l\};\tau_b) + a_1(\{w_l\};\tau_b) \cdot X(w;\tau_b)\\&
	+ a_2(\{w_l\};\tau_b)\cdot Y(w;\tau_b) + a_3(\{w_l\};\tau_b) \cdot \Big(X(w;\tau_b)^2 - \tfrac{1}{18}E_4(\tau_b)\Big)\nonumber
\end{align}
where the coefficients are given by 
(with $\theta_n \equiv \theta_n(0;\tau_b)$ being understood)
\begin{align}
	a_3(\{w_l\};\tau_b) =& \frac{1}{16}\prod_{l=1}^4\frac{\theta_1(w_l;\tau_b)}{\eta(\tau_b)^3}, \quad a_2(\{w_l\};\tau_b) = -\frac{1}{4} \prod_{l<m}^3\frac{\theta_1(w_l +w_m;\tau_b)}{\eta(\tau_b)^3}\\
	a_1(\{w_l\};\tau_b) =& -\frac{\theta_2(0;\tau_b)^4}{8}\frac{\prod_{i=1}^4\theta_3(w_i;\tau_b)+\prod_{i=1}^4\theta_4(w_i;\tau_b)}{\eta(\tau_b)^{12}}\\&
	- \frac{\theta_3(0;\tau_b)^4+\theta_4(0;\tau_b)^4}{24}\frac{\prod_{i=1}^4\theta_1(w_i;\tau_b)-3\prod_{i=1}^4\theta_2(w_i;\tau_b)}{\eta(\tau_b)^{12}} \nonumber \\
	a_0(\{w_l\};\tau_b)  =& -\frac{4\theta_2^8-3\theta_3^8+3\theta_4^8}{288}\frac{\prod_{i=1}^4\theta_3(w_i;\tau_b)}{\eta(\tau_b)^{12}}
	+ \frac{4\theta_2^8+3\theta_3^8-3\theta_4^8}{288}\frac{\prod_{i=1}^4\theta_4(w_i;\tau_b)}{\eta(\tau_b)^{12}} \\&
	- \frac{2\theta_2^8+\theta_3^8+\theta_4^8}{288}\frac{\prod_{i=1}^4\theta_1(w_i;\tau_b)}{\eta(\tau_b)^{12}} 
	- \frac{2\theta_2^8-3\theta_3^8-3\theta_4^8}{288}\frac{\prod_{i=1}^4\theta_2(w_i;\tau_b)}{\eta(\tau_b)^{12}}. \nonumber
\end{align}

\paragraph{SO(8)} 

We notice that the section $s^{B} (w,\{w_l\};\tau_f) $ of the $SO(8)$ determinant line bundle can be regarded as the product of four $SL(2)$ sections, i.e.,
\begin{align}
\label{eq:SO8-decompose}
	s^{D_{4}} (w,\{w_{1,2,3,4}\};\tau_f) = \prod_{l=1}^4 s_1^A(w,\{w_l\};\tau_f) = \sum_{n=0}^4 a_n(\{w_{l}\};\tau_f) \cdot X(w;\tau_f)^n
\end{align}
where the coefficients are given by
\begin{align}
\label{eq:SO8-coeff}
	a_0(\{w_l\};\tau_f) &= \prod_{l=1}^4\frac{\theta_1(w_l;\tau_f )^2 \wp (w_l;\tau_f)}{\eta(\tau_f)^6 }, \quad a_2(\{w_l\};\tau_f) = \prod_{l=1}^4\frac{\theta_1(w_l;\tau_f )^2}{16\eta(\tau_f)^6 } \cdot \sum_{i \neq j}   \wp (w_i;\tau_f)\wp (w_j;\tau_f), \nonumber\\
	a_1(\{w_l\};\tau_f) &= -\prod_{l=1}^4\frac{\theta_1(w_l;\tau )^2}{4\eta(\tau_f)^6 } \cdot \sum_{i \neq j\neq k \neq i}   \wp (w_i;\tau_f)\wp (w_j;\tau_f) \wp (w_k;\tau_f)\\
	a_3(\{w_l\};\tau_f) &= -\prod_{l=1}^4\frac{\theta_1(w_l;\tau_f )^2}{64\eta(\tau_f)^6 } \cdot \sum_{i }   \wp (w_i;\tau_f), \quad
	a_4(\{w_l\};\tau_f) =  \prod_{l=1}^4\frac{\theta_1(w_l;\tau )^2}{256\eta(\tau_f)^6 }. \nonumber
\end{align}
We want to express these coefficients \eqref{eq:SO8-coeff} using five $SO(8)$ fundamental characters. Four of them are level-1, associated to the irreducible representations whose highest weights are
\begin{align}
\mathbf{g}^{[0]} = [0,0,0,0],\quad\mathbf{g}^{[1]} = [1,0,0,0],\quad \mathbf{g}^{[3]} = [\tfrac{1}{2},\tfrac{1}{2},\tfrac{1}{2},\tfrac{1}{2}],\quad \mathbf{g}^{[4]} = [\tfrac{1}{2},\tfrac{1}{2},\tfrac{1}{2},-\tfrac{1}{2}]
\end{align}
in the orthogonal basis. One can explicitly write them as follows:
\begin{align}
\label{eq:so8-theta}
\Theta^{[0]}_{D_4}  (\{w_l\}; \tau_f) &= \frac{\prod_{i=1}^4 \theta_3(w_i;\tau_f) + \prod_{i=1}^4 \theta_4(w_i;\tau_f)}{2}, \\
\Theta^{[1]}_{D_4}  (\{w_l\}; \tau_f)  &= \frac{\prod_{i=1}^4 \theta_3(w_i;\tau_f) - \prod_{i=1}^4 \theta_4(w_i;\tau_f)}{2}, \nonumber\\
 \Theta^{[3]}_{D_4}  (\{w_l\}; \tau_f)  &= \frac{\prod_{i=1}^4 \theta_2(w_i;\tau_f) + \prod_{i=1}^4 \theta_1(w_i;\tau_f)}{2} \nonumber\\
\Theta^{[4]}_{D_4}  (\{w_l\}; \tau_f)  &= \frac{\prod_{i=1}^4 \theta_2(w_i;\tau_f) -  \prod_{i=1}^4 \theta_1(w_i;\tau_f)}{2}.\nonumber
\end{align}
The level-2 fundamental character is for the irreducible representation whose highest weight is $\mathbf{g}^{[2]}= [1,1,0,0]$. Some $SO(8)$ level-2 characters  can be constructed from the $SU(8)$ level-1 theta function based on the  $SO(8) \subset SU(8)$ embedding, i.e.,  ($q_f \equiv e^{2\pi i\tau_f}, \, t_i \equiv e^{2\pi i w_i}$)
\begin{align}
	\label{eq:so8-char}
	\Xi_{i}(\{w_l\};\tau_f) \equiv \textstyle\sum_{\mathbf{n},\mathbf{m} \in \mathbf{Z}^4}^{|\mathbf{m}| = i} \Big(q_f^{\mathbf{n}\cdot(\mathbf{n}-\mathbf{m}) + \frac{1}{2}\mathbf{m}^2} \cdot \prod_{i=1}^4t_i^{2n_i - m_i}\Big) .
\end{align}
For example, squares of the $SO(8)$ level-1 fundamental characters are related to them as
\begin{align}
\label{eq:so8-level1-square}
	(\Theta^{[0]}_{D_4}  (\{w_l\}; \tau_f) + \Theta^{[1]}_{D_4}  (\{w_l\}; \tau_f))^2&=\Xi_0 +2 \textstyle\sum_{i=1}^8    \Xi_i\\
	(\Theta^{[0]}_{D_4}  (\{w_l\}; \tau_f) - \Theta^{[1]}_{D_4}  (\{w_l\}; \tau_f))^2&=\Xi_0 +2 \textstyle\sum_{i=1}^8 (-1)^i\,\Xi_i\nonumber \\
	(\Theta^{[3]}_{D_4}  (\{w_l\}; \tau_f) + \Theta^{[4]}_{D_4}  (\{w_l\}; \tau_f))^2
	&=   q^{-1} \Xi_{-4} + 2 \textstyle \sum_{i=1}^8 q^{\frac{i-2}{2}}\,\Xi_{i-4} 
\nonumber \\
	(\Theta^{[3]}_{D_4}  (\{w_l\}; \tau_f) - \Theta^{[4]}_{D_4}  (\{w_l\}; \tau_f))^2&= q^{-1} \Xi_{-4} + 2 \textstyle \sum_{i=1}^8 (-1)^i q^{\frac{i-2}{2}}\,\Xi_{i-4}.
	\nonumber 
\end{align}
Expressing the coefficients \eqref{eq:SO8-coeff} using the level-1 fundamental characters and $a_3(\{w_l\};\tau_f)$, we get
\begin{align}
	a_4(\{w_l\};\tau_f) =&  \tfrac{1}{256}\,\eta(\tau_f)^{-24}\big(\Theta^{[3]}_{D_4}(\{w_l\};\tau_f) - \Theta^{[4]}_{D_4}(\{w_l\};\tau_f) \big)^2\\
	a_2(\{w_l\};\tau_f) =& +\frac{\theta_2^8 \cdot  \big(\Theta^{[0]}_{D_4}(\{w_l\};\tau_f) - \Theta^{[1]}_{D_4}(\{w_l\};\tau_f)\big)^2}{256 \eta(\tau_f)^{24}} - \frac{\theta_2^4\theta_4^4 \cdot  \Theta^{[0]}_{D_4}(\{w_l\};\tau_f) \Theta^{[1]}_{D_4}(\{w_l\};\tau_f)}{64 \eta(\tau_f)^{24}} \nonumber\\
	&- \frac{(2\theta_3^4\theta_4^4-\theta_2^8) \cdot  \big(\Theta^{[3]}_{D_4}(\{w_l\};\tau_f) - \Theta^{[4]}_{D_4}(\{w_l\};\tau_f)\big)^2}{768 \eta(\tau_f)^{24}}	\nonumber\\&
	+ \frac{\theta_3^4\theta_4^4 \cdot  \Theta^{[3]}_{D_4}(\{w_l\};\tau_f) \Theta^{[4]}_{D_4}(\{w_l\};\tau_f)}{64 \eta(\tau_f)^{24}}\nonumber\\
	a_1(\{w_l\};\tau_f)  =& -\frac{E_4(\tau_f) a_3(\{w_l\};\tau_f)}{3} + \frac{\theta_2^4\theta_3^4  \left( \theta_2^4+ \theta_3^4\right)\, \Theta^{[0]}_{D_4}(\{w_l\};\tau_f)\Theta^{[1]}_{D_4}(\{w_l\};\tau_f) }{192\eta(\tau_f)^{24}} 
	\nonumber\\&
	+ \frac{\theta_3^4 \theta_4^4    \left(\theta_3^4+\theta_4^4\right) \,\Theta^{[3]}_{D_4}(\{w_l\};\tau_f)\Theta^{[4]}_{D_4}(\{w_l\};\tau_f)}{192 \eta(\tau_f)^{24}}	\nonumber\\&
   + \frac{\left(\theta_3^4+\theta_4^4\right)^3 \big(\Theta^{[3]}_{D_4}(\{w_l\};\tau_f) - \Theta^{[4]}_{D_4}(\{w_l\};\tau_f)\big)^2}{3456 \eta(\tau_f)^{24}}\nonumber\\&
   -\frac{\theta_2^8\left(\theta_3^4+\theta_4^4\right)
   \big(\Theta^{[0]}_{D_4}(\{w_l\};\tau_f) - \Theta^{[1]}_{D_4}(\{w_l\};\tau_f)\big)^2}{384 \eta(\tau_f)^{24}} \nonumber\\
   	a_0(\{w_l\};\tau_f)  =& -\frac{2E_6(\tau_f) a_3(\{w_l\};\tau_f)}{27}   -\frac{\theta_3^4 \theta_4^4 \left(\theta_2^4+\theta_3^4\right)
   \left(\theta_2^4-\theta_4^4\right) (\Theta^{[3]}_{D_4}(\{w_l\};\tau_f)+\Theta^{[4]}_{D_4}(\{w_l\};\tau_f))^2}{2304 \eta(\tau_f)^{24}} \nonumber\\&
   +\frac{\theta_2^2 \theta_4^4 \Theta^{[0]}_{D_4}(\{w_l\};\tau_f) \Theta^{[1]}_{D_4}(\{w_l\};\tau_f) \left(\theta_2^4+\theta_3^4\right)
   \left(\theta_3^4+\theta_4^4\right)}{576 \eta(\tau_f)^{24}} \nonumber\\&+\frac{\left(\theta_3^8-\theta_4^8\right)^2 (\Theta^{[0]}_{D_4}(\{w_l\};\tau_f)-\Theta^{[1]}_{D_4}(\{w_l\};\tau_f))^2}{2304 \eta(\tau_f)^{24}} .\nonumber
\end{align}
We remark that the coefficient $ a_3(\{w_l\};\tau_f)$ arises in front of the combination $X(w;\tau_f)^3 - \frac{1}{3}E_4(\tau_f)X(w;\tau_f) - \frac{2}{27}E_6(\tau_f)$, which becomes $\frac{1}{4}Y(w;\tau_f)^2$ by the Weierstrass equation  \eqref{eq:Weierstrass-eq}. The coefficient $a_3(\{w_l\};\tau_f)$ itself can also be expressed as
\begin{align}
	a_3 (\{w_l\};\tau_f) =& +\frac{\theta_3^2 \theta_4^2}{16\eta(\tau_f)^{24}}\cdot (\Xi_2(\{w_l\};\tau_f) +\Xi_6(\{w_l\};\tau_f))	\nonumber\\&
	-\frac{\theta_4^2(\theta_3^2-\theta_4^2)}{256\eta(\tau_f)^{24}}\cdot (\Theta^{[0]}_{D_4}  (\{w_l\}; \tau_f) + \Theta^{[1]}_{D_4}  (\{w_l\}; \tau_f))^2 \nonumber\\&
	+\frac{\theta_3^2(\theta_3^2-\theta_4^2)}{256\eta(\tau_f)^{24}} \cdot (\Theta^{[0]}_{D_4}  (\{w_l\}; \tau_f) - \Theta^{[1]}_{D_4}  (\{w_l\}; \tau_f))^2 	\nonumber\\&
	-\frac{\theta_3^2\theta_4^2}{256\eta(\tau_f)^{24}} \cdot (\Theta^{[3]}_{D_4}  (\{w_l\}; \tau_f) + \Theta^{[4]}_{D_4}  (\{w_l\}; \tau_f))^2\nonumber\\
   &+\frac{\theta_3^4-3\theta_3^2\theta_4^2+\theta_4^4}{768\eta(\tau_f)^{24}}\cdot (\Theta^{[3]}_{D_4}  (\{w_l\}; \tau_f) - \Theta^{[4]}_{D_4}  (\{w_l\}; \tau_f))^2.
 	\end{align}
 This contains the level-2 characters $ \Xi_2(\{w_l\};\tau_f)$ and $\Xi_6(\{w_l\};\tau_f) = q_b^2\cdot  \Xi_2(\{w_l\};\tau_f)$ which are independent from the ones appearing in \eqref{eq:so8-level1-square}.

\subsection{Fiber-base duality}
\label{sec:fibersections}

Let us construct the Seiberg-Witten curves for 6d $\mathcal{N}=(1,0)$ little string theories $\mathcal{T}^A_{D_4,N}$ and $\mathcal{T}^B_{D_4,N}$. We begin by focusing on the case $N=1$. We have already obtained the equations for these curves as restrictions to $\mathbb{T}_{\tau_b}$ (\ref{eq:SU2-decompose}) and to $\widehat{\mathbb{T}}_{\tau_f}$ (\ref{eq:SO8-decompose}). As we have seen, these equations admit expansions in terms of $\widehat{A}_1$-characters in the one case and $\widehat{D}_4$ characters in the other. These characters are invariant under the Weyl reflections (\ref{eq:iWeyl}) and are thus sections of line bundles on the orthogonal elliptic curve in each case. In this Section, we want to find out which sections they correspond to and this way lift the expressions for the spectral curves to the four-torus $\mathbb{T}_{\tau_b} \times \widehat{\mathbb{T}}_{\tau_f}$. Our starting point is the spectral curve equation for affine $\hat{A}_1$ base. By using equations (\ref{eq:identity})--(\ref{eq:theta4ch}), we rewrite (\ref{eq:SU2-decompose}) as follows
\begin{align}
\label{eq:basecurve-sosp}
	0=  \frac{\theta_3(0;\tau_b)^{2}\theta_2(0;\tau_b)^2}{4\eta(\tau_b)^6}\cdot \theta_4(w_1;\tau_b)^2 -\left( {\theta_3(0;\tau_b)^4 + \theta_2(0;\tau_b)^4\over 12 } + \frac{X (w;\tau_b)}{4}\right) \cdot \frac{\theta_1(w_1;\tau_b)^2}{\eta(\tau_b)^6}.
\end{align}
We see that $\theta_4(w_1(z);\tau_b)^2$ and $\theta_1(w_1(z);\tau)b)^2$ are both linear combinations of $\widehat{A}_1$ characters and are thus invariant under crossing branch cuts on the $z$-plane. We thus have to replace them with sections of powers of the canonical bundle $K$ of $\widehat{\mathbb{T}}_{\tau_f}$. Here we will simply state the replacement rule and give a derivation in Section \ref{sec:SCFT} where we will be proving that the choices we make here are in fact the unique ones giving the correct SCFT limit of the LST. 
We replace the combinations of $SU(2)$ characters by the following $Sp(0)$ and $SO(8)$ fiber sections:
\begin{align}
	\label{eq:master-sosp}
	\frac{\theta_1(w_1;\tau_b)^2}{\eta(\tau_b)^6} \rightarrow&\ c_0\, \frac{\eta(\tau_f)^{18} \theta_1 (2z;\tau_f)^2}{\theta_1(z;\tau_f)^8}
	 = \frac{c_0}{64} \cdot Y(z;\tau_f)^2\\
 	\theta_4(w_1;\tau_b)^2 \rightarrow&\ c_1 \, \prod_{l=1}^4 \frac{\theta_1(z+z_l; \tau_f)\theta_1(z-z_l; \tau_f)}{\theta_1(z;\tau_f)^2}=  
 	c_1 \sum_{n=0}^4 a_n(\mathbf{z}; \tau_f) \cdot X(z;\tau_f)^n\nonumber
\end{align}
At this point we observe that this choice is consistent with the $\mathbb{Z}_2$-symmetry of (\ref{eq:Z2}). In fact, $Y(z;\tau_f)$ is an odd function under $z \mapsto -z$ as it should be because under the $\mathbb{Z}_2$ reflection we have
\begin{equation}
	\theta_1(w_1;\tau_b) \mapsto - \theta_1(w_1;\tau_b).
\end{equation}
Moreover, $\theta_4(w_1;\tau_b)$ is an even function which is again consistent with (\ref{eq:master-sosp}).
We can now proceed to use the Weierstrass equation $Y(z;\tau_f)^2 = 4X(z;\tau_f)^3 - \frac{4}{3}E_4(\tau_f)X(z;\tau_f) - \frac{8}{27}E_6(\tau_f)$, to express the Seiberg-Witten curve \eqref{eq:basecurve-sosp} as a polynomial in $X(w;\tau_b)$ and $X(z;\tau_f)$, i.e.
\begin{align}
\label{eq:curve-su2}
	0  &=   \Big( \tfrac{c_1}{4} f_0(\tau_b)a_3(\mathbf{z}; \tau_f) -\tfrac{c_0}{64}\big(\tilde{f}_0(\tau_b) + \tfrac{1}{4} X (w;\tau_b)\big) \eta(\tau_f)^{-12} \Big)Y(z;\tau_f)^2 \nonumber\\&
+ c_1 f_0(\tau_b)  \cdot\big(a_4(\mathbf{z}; \tau_f) X(z;\tau_f)^4+a_2(\mathbf{x}; \tau_f) X(z;\tau_f)^2+\tilde{a}_1(\mathbf{z}; \tau_f)X(z;\tau_f)+\tilde{a}_0(\mathbf{z}; \tau_f) \big),
\end{align}
where
\begin{align}
	f_0(\tau_b) &=  \tfrac{1}{4}\theta_3(0;\tau_b)^{2}\theta_2(0;\tau_b)^2\eta(\tau_b)^{-6}, &\tilde{f}_0(\tau_b) &=  \tfrac{1}{12} \big(\theta_3(0;\tau_b)^4 + \theta_2(0;\tau_b)^4\big)\\
	\tilde{a}_1(\mathbf{z}; \tau_f) &= a_1(\mathbf{z}; \tau_f) +\tfrac{1}{3}E_4(\tau_f)a_3(\mathbf{z}; \tau_f), & \tilde{a}_0(\mathbf{z}; \tau_f) &=a_0(\mathbf{z}; \tau_f) +\tfrac{2}{27}E_6(\tau_f)a_3(\mathbf{z}; \tau_f)
\end{align}
On the other hand, if we start from the curve equation for affine $\widehat{D}_4$ base of $\mathcal{T}^B_{D_4,1}$ and apply the following master equations derived in \cite{Nekrasov:2012xe,Haghighat:2017vch}, which replace the level-2 characters by the $SU(1)$ and $SU(2)$ fiber sections:
\begin{gather}
	(\Theta^{[0]}_{D_4}  (\mathbf{w}; \tau_f) + \Theta^{[1]}_{D_4}  (\mathbf{w}; \tau_f))^2 \rightarrow\ b_0, \qquad
	(\Theta^{[0]}_{D_4}  (\mathbf{w}; \tau_f) - \Theta^{[1]}_{D_4}  (\mathbf{w}; \tau_f))^2 \rightarrow\ b_1, \\
	\frac{(\Theta^{[4]}_{D_4}  (\mathbf{w}; \tau_f) - \Theta^{[3]}_{D_4}  (\mathbf{w}; \tau_f))^2}{\eta(\tau_f)^{24}}\rightarrow\ b_3, \qquad
	\frac{(\Theta^{[4]}_{D_4}  (\mathbf{w}; \tau_f) + \Theta^{[3]}_{D_4}  (\mathbf{w}; \tau_f))^2}{\eta(\tau_f)^{24}}\rightarrow\ b_4,
\end{gather}
as well as
\begin{align}
		\frac{\Xi_2(\mathbf{w}; \tau_f) + \Xi_6(\mathbf{w}; \tau_f)}{\eta(\tau_f)^{24}}\rightarrow&\  b_2\,  \bigg(f_0(\tau_b)\,\theta_4(z_1;\tau_b)^2 -\left(\tilde{f}_0 (\tau_b)  + \frac{X (z;\tau_b)}{4}\right) \cdot \frac{\theta_1(z_1;\tau_b)^2}{\eta(\tau_b)^6}\bigg),
\end{align}
the curve equation can also be expressed as a polynomial in $X(t;\tau_b)$ and $X(z;\tau_f)$, i.e.
\begin{align}
	0 &=\tfrac{b_3}{256}  \, X(w;\tau_f)^4 +\textstyle\sum_{i=0}^3 g_{i}(b_0,b_1,b_3,b_4;\tau_f)\, X(w;\tau_f)^i 	+ \tfrac{b_2}{16} \theta_3(0;\tau_f)^4\theta_4(0;\tau_f)^2 Y(w;\tau_f)^2 \,  \nonumber\\&\times
\left(f_0(\tau_f)\theta_4(z_1;\tau_b)^2 -(\tilde{f}_0 (\tau_b)+\tfrac{1}{4}X (z;\tau_b))\eta(\tau_b)^{-6}\theta_1(z_1;\tau_b)^2 \right).
\end{align}
We notice that it is in the same functional form with \eqref{eq:curve-su2}. The crucial miracle which made this possible is that the level-2 characters of $\widehat{D}_4$ only appear in the $a_3(\{w_l\};\tau_f)$ coefficient in the expansion of $s^B$ and that coefficient is only multiplying $\frac{1}{4} Y(w;\tau_f)^2$!

\subsection{SCFT limit}
\label{sec:SCFT}

Let us take the 6d SCFT reduction, i.e., $R_6/\ell_s \rightarrow \infty$, which decompactifies the toroidal base $\mathbb{T}_{\tau_b}$ into a cylindrical one by sending $\tau_b  \rightarrow i\infty$, or equivalently, $q_b \equiv q_0 q_1 \rightarrow 0$. There are two options for decompactification. The first option is to take $q_0 \rightarrow 0$, sending the 2d $O(2k)$ gauge coupling to zero. The resulting SCFT is the 6d $SO(8)$-theory with zero hypermultiplets as studied in \cite{Haghighat:2014vxa}. In this limit, the profile functions \eqref{eq:profile-so8} get reduced to $g(z) = 0$, such that $t_1 \equiv e^{2\pi i w_1(z;\tau_f)}$ can be identified as 
\begin{align}
	t_1 = \frac{1}{\sqrt{\mathcal{P}_1(z;\tau_f)}}\, \frac{y_1(z;\tau_f)}{y_0(z;\tau_f)} \ \longrightarrow \ \frac{1}{\sqrt{q_1} \theta_1(2z;\tau_f)^2} y_1(z;\tau_f).
\end{align}
Then the curve equation \eqref{eq:basecurve-sosp} becomes (with $t \equiv e^{2\pi i w}$)
\begin{align}
	\label{eq:6dscft-so8}
	0 = \sqrt{q_1} \theta(2z;\tau_f)^2 t^2 - \chi_1(z;\tau_f) \; t + \sqrt{q_1}\theta_1(2z;\tau_f)^2 \  \text{ with } \  \chi_1 \equiv y_1 + \frac{q_1 \theta_1(2z;\tau_f)^4}{y_1}
\end{align}
where for consistency reasons $\chi_1(z;\tau_f)$ has to be a section of a degree $8$ line bundle over $\widehat{\mathbb{T}}_{\tau_f}$\footnote{Basically, the reason is that as explained in the Appendix $\theta(2z;\tau)^2$ is a degree $8$ section and therefore $\chi_1$ must also be.}.  Another 6d SCFT can be reached by taking the limit $q_1 \rightarrow 0$, i.e., 	$R_6/\ell_s \rightarrow \infty$ after $w \rightarrow w' = -\frac{\tau_b}{2} - w$ and $w_1 \rightarrow w_1' = -\frac{\tau_b}{2} - w_1$. This induces 6d rank-1 E-string theory with $E_8 \rightarrow SO(8)$ holonomy, also known as $D_4$ conformal matter theory \cite{DelZotto:2014hpa}. Now the profile function \eqref{eq:profile-Estr} becomes $f(z) = 2 \sum_{i=1}^4 (|z - \mu_l| + |z + \mu_l|)$ such that 
\begin{align}
	t_1' = \frac{\sqrt{\mathcal{P}_1(z;\tau_f)}}{\sqrt{q_0 q_1}} \frac{y_0(z;\tau_f)}{y_1(z;\tau_f)} \ \longrightarrow\  \frac{\theta_1(2z;\tau_f)^2}{\sqrt{q_0}\, \prod_{l=1}^4\theta_1(z \pm \mu_l)} y_0(z;\tau_f).
\end{align}
In this case the curve equation \eqref{eq:basecurve-sosp} becomes (with $t \equiv e^{2\pi i w}$)
\begin{align}
	\label{eq:6dscft-Estr}
	0 = \sqrt{q_0} \prod_{l=1}^4\theta_1(z \pm \mu_l)\, t^2 - \theta_1(2z;\tau_f)^2 \cdot \chi_0(z;\tau_f) \; t + \sqrt{q_0}\prod_{l=1}^4\theta_1(z \pm \mu_l)
\end{align}
where $\chi_0(z;\tau_f)$ is a section of a degree~0 line bundle, defined by
\begin{align}
\chi_0 \equiv y_0 + \frac{q_0 \prod_{l=1}^4\theta_1(z \pm \mu_l)^2}{\theta_1(2z;\tau_f)^4}\frac{1}{y_0}.
\end{align}
Now we want to derive the LST curve \eqref{eq:curve-su2}, or equivalently, \eqref{eq:basecurve-sosp} after imposing the replacement rule \eqref{eq:master-sosp} by unfreezing an ``$Sp(0)$"/$SO(8)$ node from \eqref{eq:6dscft-so8}/\eqref{eq:6dscft-Estr} respectively. We observe that under $q_b \rightarrow 0$ we have the following limiting behaviors
\begin{eqnarray}
	\frac{\theta_3(0;\tau_b)^{2}\theta_2(0;\tau_b)^2}{4\eta(\tau_b)^6} & \longrightarrow 1 & ,  \\
	{\theta_3(0;\tau_b)^4 + \theta_2(0;\tau_b)^4\over 12 } + \frac{X (w;\tau_b)}{4} & \longrightarrow & - \frac{t}{(t-1)^2}.
\end{eqnarray}
Together with the replacement rules \eqref{eq:master-sosp} the limiting behavior of equation \eqref{eq:basecurve-sosp} is then
\begin{align}
	\label{eq:6dscft-reduction-1}
	0 = c_1 \, \frac{\prod_{l=1}^4\theta_1(z \pm \mu_l;\tau_f)}{\theta_1(z;\tau_f)^8} (t-1)^2 + c_0 \frac{\eta(\tau_f)^{18}\theta_1 (2z;\tau_f)^2}{\theta_1(z;\tau_f)^8} t
\end{align}
Here we have used the fact that the variables $z_l$ are functions of coupling constants, fugacities and other scales in the theory such that under $q_{b,1} \rightarrow 0$ we have
\begin{equation}
	z_l(\tau_{b,1},\tau_f,\{\mu_l\}) \longrightarrow \mu_l.
\end{equation}
We see that \eqref{eq:6dscft-reduction-1} can be immediately put into the form \eqref{eq:6dscft-Estr} with the identification
\begin{equation} \label{eq:chi0}
	\chi_0(z;\tau_f) = \frac{2 \sqrt{q_0}\prod_{l=1}^4 \theta_1(z\pm \mu_l;\tau_f)}{\theta_1(2z;\tau_f)^2} - \frac{c_0}{c_1} \sqrt{q_0}\,\eta(\tau_f)^{18} . 
\end{equation}
Another limit we can take is obtained by first shifting $w \rightarrow w' = -\frac{\tau_b}{2} - w$ and $w_1 \rightarrow w_1' = -\frac{\tau_b}{2} - w_1$. Under this shift, the role of $\theta_1$ and $\theta_4$ gets exchanged by the identities 
\begin{eqnarray}
	\theta_4(-\frac{\tau_b}{2} - w_1;\tau_b) & = & q_b^{-1/8} e^{i w_1 \pi} \theta_1(w_1;\tau_b), \nonumber \\
	\theta_1(-\frac{\tau_b}{2} - w_1;\tau_b) & = & q_b^{-1/8} e^{i w_1 \pi} \theta_4(w_1;\tau_b).
\end{eqnarray}
Thus using the replacement rules\eqref{eq:master-sosp}, the limiting behavior of the spectral curve under $q_b \rightarrow 0$ becomes
\begin{align}
	\label{eq:6dscft-reduction-2}
	0 = c_1 \, \frac{\prod_{l=1}^4\theta_1(z \pm z_l;\tau_f)}{\theta_1(z;\tau_f)^8} t' + c_0 \frac{\eta(\tau_f)^{18}\theta_1 (2z;\tau_f)^2}{\theta_1(z;\tau_f)^8} (t'-1)^2, \quad t' = q_b^{-1/2}/t_b.
\end{align}
Note that here $z_l$ does not get replaced with $\mu_l$ as we are keeping $\tau_{b,1}$ finite in the limit. Again, one can easily see that equation \eqref{eq:6dscft-reduction-2} can be recast into the form \eqref{eq:6dscft-so8} with the identification
\begin{equation} \label{eq:chi1}
	\chi_1 = 2 \sqrt{q_1}\, \theta_1(2z;\tau_f)^2 - \frac{c_1}{c_0} \frac{\sqrt{q_1}\prod_{l=1}^4 \theta(z \pm z_l;\tau_f)}{\eta(\tau_f)^{18}}.
\end{equation}
Performing these steps, we have learned two things. Firstly, the replacement rules \eqref{eq:master-sosp} are the unique choices compatible with the SCFT curves \eqref{eq:6dscft-so8} and \eqref{eq:6dscft-Estr}, in particular they reproduce the corresponding matter polynomials in a correct way. Secondly, we have derived expressions for the SCFT characters given by equations \eqref{eq:chi0} and \eqref{eq:chi1}. Armed with these expressions, we can next proceed to take the 5d and 4d limits.

\subsubsection{5d/4d SCFTs}

The 6d LST $\mathcal{T}^A_{D_4,1}$ can be reduced to the 5d SCFT in the decompactification limit $q_f \rightarrow 0$, accompanied by $SO(8)$ holonomy $(z_1, z_2, z_3, z_4) = (z_1,z_2,\frac{\tau_f}{2}-z_3,\frac{\tau_f}{2}-z_4)$ that breaks $SO(8)$ to $SO(4)^2$. The elliptic curve $\mathbb{T}_{\tau_f}$ will be degenerated to the cylinder $\mathbb{C}_{\tau_f}^\times$. The resulting 5d SCFT will be effectively described by 5d $SO(4) \times ``Sp(0)"$ circular quiver gauge theory, whose curve equation is obtained from \eqref{eq:curve-su2} as follows:
\begin{align}
\label{eq:5dscft-sosp}
	0=  \frac{\theta_3(0;\tau_b)^{2}\theta_2(0;\tau_b)^2}{4\eta(\tau_b)^6}\cdot \prod_{i=1}^2\text{sn}\,(z\pm z_l) -\left( {\theta_3(0;\tau_b)^4 + \theta_2(0;\tau_b)^4\over 12 } + \frac{X (w;\tau_b)}{4}\right) \cdot \text{sn}\,(2z)^2	
\end{align}
where $\text{sn}(x) \equiv 2\sin{(\pi x)}$. 
In the dual description $\mathcal{T}^B_{D_4,1}$, the same decompactification limit corresponds to freezing two ``$SU(1)$" nodes, yielding the linear $``SU(1)" \times SU(2) \times ``SU(1)"$ quiver gauge theory with two hypermultiplets attached to the middle $``SU(2)"$ node  \cite{Hayashi:2015vhy}. 

Extra decompactification of the $X^1$ circle removes periodicity $z \sim z + 1$ from a fiber coordinate $z$, turning the cylinder $\mathbb{C}_{\tau_f}^\times$ to the complex plane $\mathbb{C}_{\tau_f}$. Such a limit will reduce \eqref{eq:5dscft-sosp} to a polynomial equation in fiber coordinates, corresponding to the 4d spectral curve. To reach the conformal gauge theory whose $\beta$ function vanishes to zero, it is also required to introduce the $SO(4)$ holonomy $(z_1, z_2) = (z_1, \frac{1}{2}-z_2)$ before decompactification, breaking $SO(4) \rightarrow SO(2)^2$. The resulting 4d SCFT will have $SO(2) \times ``Sp(0)"$ gauge symmetry and the following curve equation:
\begin{align}
\label{eq:4dscft-sosp}
	0=  \frac{\theta_3(0;\tau_b)^{2}\theta_2(0;\tau_b)^2}{4\eta(\tau_b)^6}\cdot (z^2 - z_1^2) -\left( {\theta_3(0;\tau_b)^4 + \theta_2(0;\tau_b)^4\over 12 } + \frac{X (w;\tau_b)}{4}\right) \cdot 4z^2.
\end{align}
This is consistent with the analysis of \cite{Landsteiner:1997vd}; especially the polynomial degree in $z$ matches.

Here we also apply the same 5d/4d reductions directly to 6d SCFT curves. Firstly, 6d $D_4$ conformal matter becomes 5d $\mathcal{N}=1$ free QFT of four $\tfrac{1}{2}$-hypermultiplets with $SO(4)$ flavor symmetry. Its curve equation can be derived from \eqref{eq:6dscft-reduction-1} as follows:
\begin{align}
	\label{eq:5dscft-reduction-1}
\textstyle	0 = c_1 \, \prod_{i=1}^2\text{sn}\,(z\pm \mu_l)\cdot   (t-1)^2 + c_0 \,\text{sn}\,(2z)^2\cdot t
\end{align}
The 4d reduction gives the free QFT of two $\tfrac{1}{2}$-hypers with $SO(2)$ flavor group, for which
\begin{align}
\textstyle	\label{eq:4dscft-reduction-1}
	0 = c_1 \, (z^2 - \mu_1^2)\cdot   (t-1)^2 + c_0 \,(2z)^2\cdot t.
\end{align}
Secondly, 6d $SO(8)$ SCFT reduces to 5d $\mathcal{N}=1$ $SO(4)$ super Yang-Mills, whose curve is
\begin{align}
\textstyle	\label{eq:5dscft-reduction-2}
	0 = c_0 \, \text{sn}\,(2z)^2\cdot   (t-1)^2 + c_1  \, \prod_{i=1}^2\text{sn}\,(z\pm z_l)\cdot t.
\end{align}
Further reduction will yield 4d $\mathcal{N}=2$ $SO(2) \simeq U(1)$ gauge theory, for which 
\begin{align}
\textstyle	\label{eq:4dscft-reduction-2}
	0 = c_0 \, (2z)^2\cdot   (t-1)^2 + c_1  \, (z^2 - z_1^2)\cdot t.
\end{align}
Notice that the 4d curve equations \eqref{eq:4dscft-reduction-1} and \eqref{eq:4dscft-reduction-2} are all in agreement with \cite{Landsteiner:1997vd}.

\subsection{Some comments on $N=2$}

Let us give some comments on the case of $2$ M5 branes probing the $D_4$ singularity corresponding to the theory $\mathcal{T}^A_{D_4,2}$. The restriction of its spectral curve to $\mathbb{T}_{\tau_b}$ is now the determinant section of an $SL(4)$ bundle given in equation \eqref{eq:sl4}. The SW-curve of the dual theory is obtained by first writing down the section $s^{D_4}$ and then subsequently replacing the level-2 $D_4$ characters by $SL(4)$ sections. Looking at \eqref{eq:sl4}, we see that the highest power of $X(w;\tau_b)$ is quadratic and multiplies the iWeyl-invariant coefficient $a_3(\{w_l\};\tau_b)$. Thus in order for fiber-base duality to work, we have to introduce the replacement rule
\begin{equation}
	a_3(\{w_l\};\tau_b) \longrightarrow c_3 Y(z;\tau_f)^2,
\end{equation}
to reproduce the structure appearing in \eqref{eq:sB}. Moreover, we see that $a_0(\{w_l\},\tau_b)$ has to be replaced by the determinant of an $SO(8)$ bundle, i.e.
\begin{equation}
	a_0(\{w_l\};\tau_b) \longrightarrow c_0 \frac{\prod_{l=1}^4\theta_1(z \pm z_l;\tau_f)}{\theta_1(z;\tau_f)^8}.
\end{equation}
The replacement rule for $a_1(\{w_l\};\tau_b)$ is expected to be a linear combination of the above two sections, i.e.
\begin{equation}
	a_1(\{w_l\};\tau_b) \longrightarrow c_1 \frac{\prod_{l=1}^4\theta_1(z \pm z_l;\tau_f)}{\theta_1(z;\tau_f)^8} + \widetilde{c}_1 Y(z;\tau_f)^2.
\end{equation}
Regarding $a_2(\{w_l\};\tau_b)$, the situation is more complicated. Naively, one might think that one should replace it with $Y(z;\tau_f)^2$ in order to reproduce the structure of the dual curve. However, this is not allowed due to the $\mathbb{Z}_2$ symmetry enjoyed by the theory $\mathcal{T}^A$. The reason is that $Y(w;\tau_b)$ is an odd function under $w \mapsto - w$, which means that $a_2$ should be replaced by an odd function under $z \mapsto -z$. Therefore, $Y(z;\tau_f)^2$ is ruled out by this. 

If that is the case, how can then the two curves match? One answer is that they will not unless one sets certain parameters in both equations to zero. In the $\mathcal{T}^A$ curve equation we would set $a_2(\{w_l\};\tau_b) = 0$ and in the $\mathcal{T}^B$ curve equation we would need to set to zero the analogous term appearing after applying the replacement rules. The interpretation of this in the F-theory geometric engineering picture, is that one has to blow-down $-1$-curves in the base of one theory and half of the $-2$-curves in the central fiber of the other theory. In fact, blowing down $-1$-curves in the base leads to a chain of $-2$-curves which can then be matched with the central fiber $-2$-curves of the other theory. A further analysis of these issues is beyond the scope of this work and we leave it for future investigations.

\section{Discussion}

In this paper we mainly studied the Seiberg-Witten curves for 6d little string theories, engineered from M5-branes probing $D_n$ singularities. This brane configuration admits a dual description in terms of D5-branes probing $D_n$ singularities, whose effective description is given by a 6d affine $\hat{D}_n$ quiver gauge theory with $U(2N) \times U(N)^4$ gauge symmetry. The curve we found is identical to the one obtained from the dual description. The curve is also reducible to the spectral curves of 6d/5d/4d SCFTs, reproducing some known results given in \cite{Landsteiner:1997vd}.

There are two interesting directions to extend the current work. One direction is to apply the same approximation techniques used in this paper and \cite{Nekrasov:2012xe,Haghighat:2017vch} to various 6d $\mathcal{N}=(1,0)$ LSTs and SCFTs. One particular pair of $(1,0)$ LSTs are $E_8 \times E_8$ and $SO(32)$ heterotic little string theories, for which we naturally expect the $Sp(N)$ determinant line bundle to be supported on the base and fiber curve, respectively. Working out the heterotic LST curves will require an interesting extension of our analysis to non-simply-laced vector bundles. It will be also interesting to study the spectral curves of 6d SCFTs supported on $\mathcal{O}(-3)$ curves, such as the $SU(3)$ SCFT, the $G_2$ SCFT with $n_f = 1$, and the $SO(7)$ SCFT with $n_s = 2$, whose 2d GLSMs producing the correct elliptic genera of  instanton strings are obtained in \cite{Kim:2016foj,Kim:2018gjo} but did \emph{not} originate from brane configurations. Likewise, non-Higgsable SCFTs on clusters of 2-cycles, i.e., $(-3)(-2)(-2)$ curves, $(-2)(-3)(-2)$ curves, and $(-3)(-2)$ curves, can be studied from their 2d GLSMs \cite{Kim:2018gjo} and are expected to give an interesting curve equation. 

Another extension of the current work is to study the quantum curve which arises in the refined topological string partition function for the elliptic Calabi-Yau manifolds. In the NS-limit the curve is expected to capture information about BPS magnetic strings \cite{Haghighat:2015coa,Hohenegger:2015cba}. More generally, the $q$-deformed and $qq$-deformed Seiberg-Witten curves of 6d $D_n$ LSTs and corresponding SCFTs, are obtained from their codimension-4 half-BPS defect partition functions \cite{Nekrasov:2013xda,Nekrasov:2015wsu}. As established in \cite{Kim:2016qqs,Kimura:2017auj,Agarwal:2018tso} for the case of $(2,0)$ SCFTs, the $\mathcal{Y}$ variables will be defined as a particular collection of residues in the GLSM contour integral, constituting the entire defect partition function  $\mathcal{X}$ that corresponds to the $\Omega$-deformation of an iWeyl character.

\acknowledgments
We would like to thank Jean-Emile Bourgine, Michele Del Zotto,  Hossein Movasati and Antonio Sciarappa for valuable discussions. JK is grateful to Yau Mathematical Sciences Center, Tsinghua University, for hospitality, where this work was partly carried out.

\appendix
\section{Notation}
\label{sec:notation}
This section collects definitions of elliptic and Jacobi elliptic functions used throughout the paper. 
First, the Jacobi theta function $\theta_i(x;\tau)$ is a section of a degree~1 line bundle over an elliptic curve $\mathbb{T}_\tau$, having quasi-periodicity ($t \equiv e^{2\pi i x},\, q \equiv e^{2\pi i \tau}$)
\begin{align}
	\theta_1 (x + j_1 + j_2 \tau;\tau) &= (-1)^{j_1+j_2} q^{-j_2^2 / 2}t^{-j_2} \theta_1 (x;\tau), \\	
	\theta_2 (x + j_1 + j_2 \tau;\tau) &= (-1)^{j_1} q^{-j_2^2 / 2}t^{-j_2}\theta_2 (x;\tau)\nonumber\\
	\theta_3 (x + j_1 + j_2 \tau;\tau) &= q^{-j_2^2 / 2}t^{-j_2}\theta_3 (x;\tau)\nonumber\\
	\theta_4 (x + j_1 + j_2 \tau;\tau) &= (-1)^{j_2} q^{-j_2^2 / 2}t^{-j_2}\theta_4 (x;\tau)\nonumber.
\end{align}
They can be written in the following series expansion form: \begin{align}
	\theta_1 (x;\tau) &=i\sum_{n=0}^\infty (-1)^n\, (t^{n+\frac{1}{2}} - t^{-n-\frac{1}{2}})\cdot q^{\frac{1}{2}(n+\frac{1}{2})^2}\\
	\theta_2 (x;\tau) &=\sum_{n=0}^\infty (t^{n+\frac{1}{2}} + t^{-n-\frac{1}{2}})\cdot q^{\frac{1}{2}(n+\frac{1}{2})^2}\nonumber\\
	\theta_3 (x;\tau) &=\sum_{n=0}^\infty (t^{n} + t^{-n})\cdot q^{\frac{1}{2}n^2},\nonumber\\
	\theta_4 (x;\tau) &=\sum_{n=0}^\infty (-1)^n\,(t^{n} + t^{-n})\cdot q^{\frac{1}{2}n^2}\nonumber.
\end{align}
These theta functions are used to construct sections of line bundles having different degrees. For example, $\theta_1(2x;\tau)^2$ shows the following quasi-periodicity,
\begin{align}
	\theta_1\big(2(x + j_1 + j_2\tau);\,\tau\big)^2 = q^{-8j_2^2 / 2}t^{-8j_2}	\theta_1\big(2x ;\tau\big)^2
\end{align}
which implies that it is a section of a degree~8 line bundle.

Secondly, the Weierstrass $\wp$-function $\wp (x;\tau)$ is a section of a degree-$0$ line bundle, which can be defined in terms of Jacobi theta functions, i.e.,
 \begin{align}
 \label{eq:wp-theta}
\wp (x;\tau)&=\frac{\theta_3(0;\tau)^{2}\theta_2(0;\tau)^2}{4}{\theta_4(x;\tau)^2 \over \theta_1(x;\tau)^2}-{\theta_3(0;\tau)^4 + \theta_2(0;\tau)^4\over 12}.
 \end{align}
We often use the following notations in Section~\ref{sec:curve}:
\begin{gather}
	\wp' (x;\tau) \equiv \tfrac{1}{\pi} \partial_x \wp (x;\tau), \ \wp^{(n)} (x;\tau) \equiv (\tfrac{1}{\pi} \partial_x)^n \wp (x;\tau), \\
	 X(x;\tau) \equiv 4\wp(x;\tau), \ Y(x;\tau) \equiv 4\wp' (x;\tau).
\end{gather}
where $X(x;\tau)$ and $Y(x;\tau)$ satisfy the Weierstrass equation
\begin{align}
\label{eq:Weierstrass-eq}
	Y(w;\tau)^2 = 4X(w;\tau)^3 - \frac{4}{3}E_4(\tau)X(w;\tau) - \frac{8}{27}E_6(\tau).
\end{align}

Finally, $E_4(\tau)$, $E_6(\tau)$, and $\eta (\tau)$ are respectively the Eisenstein series of index 4 and 6 and the Dedekind eta function, defined as
\begin{align}
 E_4(\tau) &= \frac{1}{2}\sum_{i=2}^4 \theta_i(0;\tau)^8 , \\ 
 E_6(\tau) &= -\frac{1}{2} \left(3\theta_2(0;\tau)^8\sum_{i=3}^4 \theta_i(0;\tau)^4  - \sum_{i=3}^4 \theta_i(0;\tau)^{12}\right) \\
\eta (\tau) &=q^{\frac{1}{24}}\prod_{n=1}^\infty (1-q^n).
\end{align}

\bibliographystyle{utphys}
\bibliography{ref.bib}

\end{document}